%% file: main.tex
\documentclass[sigconf,authorversion]{acmart}
\usepackage{enumitem}
\usepackage{subcaption}
\usepackage{graphicx}
\usepackage{changepage}

\AtBeginDocument{%
  \providecommand\BibTeX{{%
    \normalfont B\kern-0.5em{\scshape i\kern-0.25em b}\kern-0.8em\TeX}}}

\copyrightyear{2023}
\acmYear{2023}
\setcopyright{acmlicensed}\acmConference[CHI '23]{Proceedings of the 2023 CHI Conference on Human Factors in Computing Systems}{April 23--28, 2023}{Hamburg, Germany}
\acmBooktitle{Proceedings of the 2023 CHI Conference on Human Factors in Computing Systems (CHI '23), April 23--28, 2023, Hamburg, Germany}
\acmPrice{15.00}
\acmDOI{10.1145/3544548.3581302}
\acmISBN{978-1-4503-9421-5/23/04}

\acmSubmissionID{1417}

\begin{document}

\title[ImageAssist]{ImageAssist: Tools for Enhancing Touchscreen-Based Image Exploration Systems for Blind and Low Vision Users}


\author{Vishnu Nair}
\email{nair@cs.columbia.edu}
\affiliation{%
 \institution{Columbia University}
 \city{New York}
 \state{New York}
 \country{USA}
}

\author{Hanxiu 'Hazel' Zhu}
\email{hz2653@columbia.edu}
\affiliation{%
 \institution{Columbia University}
 \city{New York}
 \state{New York}
 \country{USA}
}

\author{Brian A. Smith}
\email{brian@cs.columbia.edu}
\affiliation{%
 \institution{Columbia University}
 \city{New York}
 \state{New York}
 \country{USA}
}

\renewcommand{\shortauthors}{Nair et al.}

\begin{abstract}
  Blind and low vision (BLV) users often rely on alt text to understand what a digital image is showing. However, recent research has investigated how touch-based image exploration on touchscreens can supplement alt text. Touchscreen-based image exploration systems allow BLV users to deeply understand images while granting a strong sense of agency. Yet, prior work has found that these systems require a lot of effort to use, and little work has been done to explore these systems' bottlenecks on a deeper level and propose solutions to these issues. To address this, we present ImageAssist, a set of three tools that assist BLV users through the process of exploring images by touch --- scaffolding the exploration process. We perform a series of studies with BLV users to design and evaluate ImageAssist, and our findings reveal several implications for image exploration tools for BLV users.
\end{abstract}




\begin{teaserfigure}
    \includegraphics[width=0.85\textwidth]{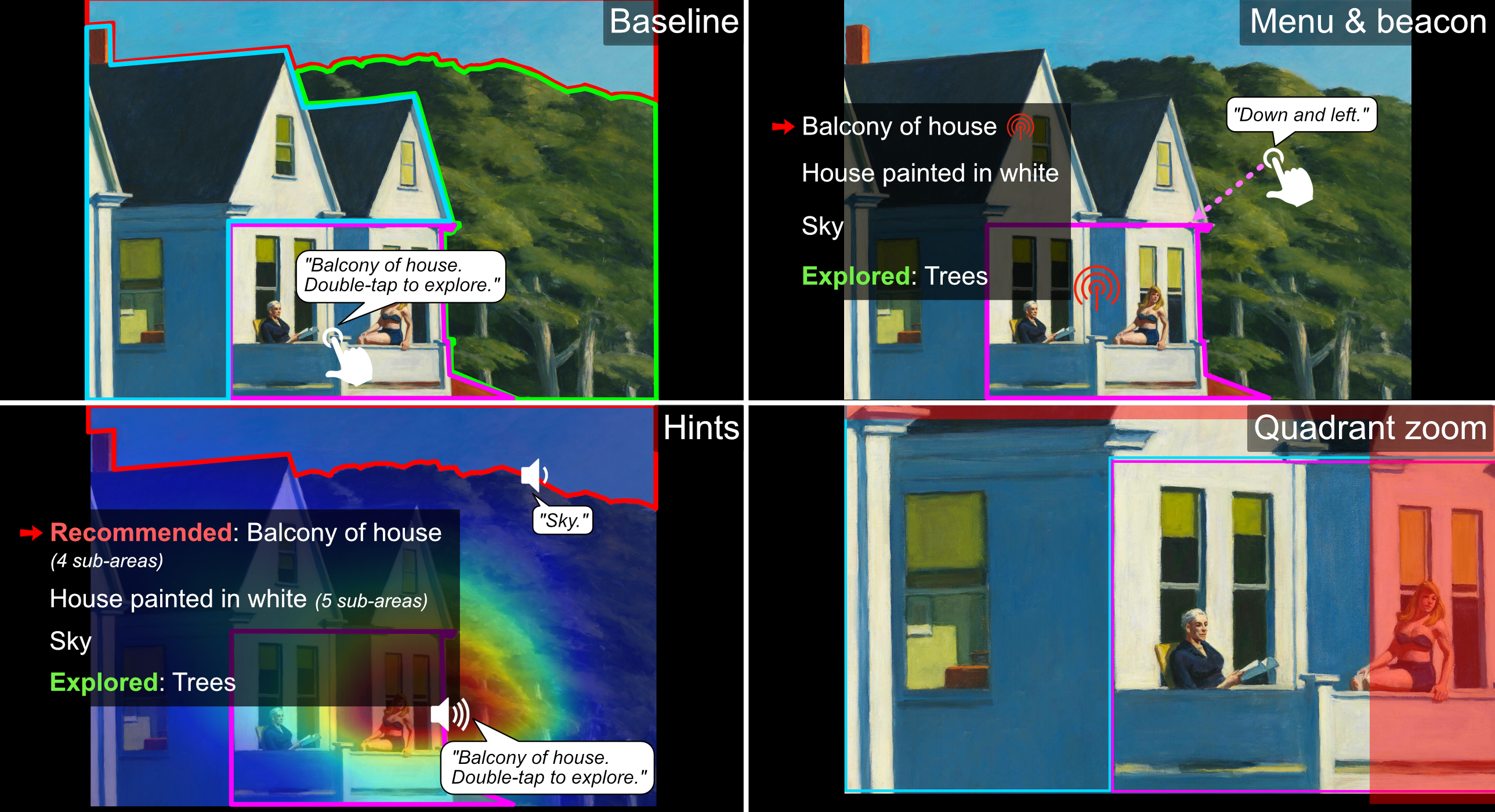}
    \centering
    \caption{Depictions of the state of the art in touchscreen-based image exploration alongside ImageAssist, a set of three tools we created to alleviate bottlenecks in this state-of-the-art/baseline system. The baseline system divides images into regions, and users touch the image to explore them. The tools that comprise ImageAssist are: A menu \& beacon tool listing areas within the image and directing users to those areas; a hints tool providing recommendations and a volume-based prominence indicator to give users a sense of direction while exploring; and a zoom tool to quickly zoom into the four quadrants of the image to survey small areas.}
    \Description{Four panel image showing the baseline touchscreen-based image exploration system plus ImageAssist's three component tools on a painting of a house alongside some trees under a blue sky.}
    \label{fig:teaser}
\end{teaserfigure}
\maketitle
\section{Introduction}
\input{sec01-intro}


\section{Related Work}
\label{section:related-work}
\input{sec02-rw}


\section{Formative Study}
\label{section:formative}
\input{sec03-formative}


\section{ImageAssist}
\label{section:probes}
\input{sec04-probes}


\section{User Study}
\label{section:us-overview}
\input{sec05-us-overview}


\section{User Study Findings}
\label{sec:us-findings}
\input{sec06-us-findings}


\section{Discussion: Implications for Touchscreen-based Image Exploration Tools}
\label{sec:discussion}
\input{sec07-discussion}


\section{Applications}
\input{sec08-applications}


\section{Limitations}
\label{sec:limitations}
\input{sec09-limitations}


\section{Conclusion}
\input{sec10-conclusion}

\begin{acks}
We would like to thank the following people: Max Tseng for his assistance during and after our main user studies; Brian (Shao-en) Ma and Jazmyn Jenkins for their help in preparing analytics scripts for our studies; and Connor Courtien for his help in preparing for our formative studies. We would also like to extend our sincere gratitude toward our study participants for their participation and to the anonymous reviewers for their helpful feedback.
\end{acks}

\bibliographystyle{ACM-Reference-Format}
\bibliography{refs}


\end{document}

%% file: sec01-intro.tex
Digital images are playing an increasingly important role in everyday life. From images uploaded to social media to digitized versions of artwork to even personal photos taken with a commodity smartphone, perceiving and understanding images are a central part of modern life. Although sighted individuals can use vision to look at images, blind and low vision (BLV) individuals are unable to do this and must rely on other methods to understand images.

BLV users often rely on alternative text (also known as ``alt text'') to understand what a digital image is showing~\cite{W3Consortium2016, McEwan2007a}. However, prior work has shown that alt text may suffer from quality issues and may be difficult to author well for complex graphics~\cite{Voykinska2016, Gleason2019, McEwan2007a, MacLeod2017, Salisbury2017, McCall2022}. This has led researchers to investigate techniques that supplement alt text, offering BLV users a richer view of images. Among these is the use of touchscreen-based devices to allow users to survey images using their finger.

Touchscreen-based image exploration systems have several advantages for BLV users over simple alt text, including granting users with a strong sense of the spatial layout of image elements, affording users with a strong sense of autonomy, and allowing BLV users to more vividly imagine what images are showing~\cite{Ahmetovic2021, Lee2022a, Morris2018}. Yet, exploring images using these systems is a process requiring far more effort than simply hearing a description: Users must move their finger around the screen to look for elements on the image while trying to make sure that they have seen everything. Indeed, prior work has identified the effort required as a major issue facing these systems~\cite{Lee2022a}. Little work, however, has been done to explore the specific interaction bottlenecks faced by BLV users in the image exploration process and propose potential solutions to alleviate these bottlenecks.

Designing solutions to the bottlenecks and pain points inherent in touchscreen-based image exploration systems is an important step to making them more usable and, thus, more accessible to BLV users --- allowing them to reap the benefits of these systems with minimal effort and frustration. Thus, in this work, we explore these pain points, and design and test tools intended to alleviate these issues. 

We first implemented a smartphone-based, touchscreen image exploration system --- meant to faithfully recreate state-of-the-art methods~\cite{Lee2022a, Microsoft2017, Morris2018, Ahmetovic2021} --- and had two BLV users try the tool out on a variety of images while providing feedback as part of a formative study. We found that participants scanned the image in an inefficient manner, and they often missed elements of the image even when they were trying to find everything. Furthermore, they experienced difficulties surveying small and tightly clustered areas of the image.

In response to these issues, we created ImageAssist, a set of three tools for making touchscreen-based image exploration more streamlined. ImageAssist comprises the following tools: 

\begin{itemize}
    \item a \textbf{menu \& beacon tool}, which presents a menu that lists out areas within the image and a beacon that directs users to areas of interest so that they do not miss elements within the image;
    \item a \textbf{hints tool}, which provides ``hints'' to give users a sense of direction while exploring the image and allow them to explore it more efficiently; and
    \item a \textbf{quadrant zoom tool}, which allows users to quickly zoom into any of the four quadrants of the image in order to survey small and tightly clustered areas more easily.
\end{itemize}

These tools are illustrated in Figure~\ref{fig:teaser}, and they supplement existing image exploration systems in an effort to alleviate the issues we discovered during our formative study.

We then presented both ImageAssist and our recreation of state-of-the-art approaches (representing the baseline condition) to nine BLV users as part of our main user study. Our evaluation showed that participants extensively used the menu in order to get an overview of the image --- especially at the very beginning and very end of exploration --- and used the beacon system as their primary tool for finding areas they were not able to find in their initial passthroughs of the image. Participants' sentiments also showed that the hints gave them a sense of direction while exploring the image and allowed them to interpret images better. Finally, while some participants felt that they were better able to explore small, tightly clustered elements within the image using the quadrant zoom tool, others thought that it was not very useful without the other tools we created to supplement it.

Our findings reveal several implications for future touchscreen-based image exploration tools, especially with respect to the importance of overviews in exploring images and the tension inherent in balancing ease-of-use with a sense of agency in accessibility tools. 

%% file: sec02-rw.tex
Our work builds upon a foundation of prior research on communicating the contents of images to BLV users.

\subsection{Presenting images to BLV users via alt text and captions.}
\label{subsec:rw-alttext}

Image descriptions, especially in the form of alt text, are the primary method by which BLV users learn about what an image is showing~\cite{W3Consortium2016, McEwan2007a}. These descriptions may be authored by humans (ideally using established guidelines~\cite{W3Consortium2016, WebAIM2021}) or generated via automated techniques~\cite{He2017, Liu2017, Ramnath2014, Wu2017} --- the latter being used extensively in mainstream tools such as Facebook’s Automatic Alt Text feature~\cite{Libaw2018, Wu2017}.

Prior research, however, has shown that alt text authored by humans may suffer from quality issues~\cite{Voykinska2016, Gleason2019, McEwan2007a}, and that auto-generated descriptions may be error-prone and misleading~\cite{MacLeod2017, Salisbury2017}. Furthermore, alt text may not be able to provide a complete overview of an image when used for complex graphics~\cite{McCall2022}.

Some researchers have focused on presenting methods and frameworks~\cite{Stangl2020, Stangl2021, Bennett2021, Mack2021, Gleason2019a} for authoring higher-quality alt text as well as on improving the quality of auto-generated descriptions~\cite{Gleason2020, Salisbury2017, Liu2017, Zhou2019}. Our work, however, focuses not on improving the quality of alt text, but on helping BLV users explore images via other methods --- in this case, via touch --- to increase their understanding of images \textit{beyond} captions.

\subsection{Presenting images to BLV users via touch-based methods.}
\label{subsec:rw-touch}

An alternative to learning about images involves exploring them via touch. Physical tactile representations have often been used to "display" images, such as diagrams and maps, to BLV users~\cite{Williams2014a}; these are often created via braille embossing~\cite{Gardner1996, Rowell2005} or 3D printing~\cite{Holloway2018, Shi2016} and may even involve the use of specially-made tactile displays, such as the Dot Pad~\cite{Coldeway2022}. However, physical tactile image representations often suffer from limited availability, high costs to users, and usability issues~\cite{Gupta2017, Rowell2005}. As such, much recent research has focused on using touchscreens --- such as those found on smartphones and tablets, and which are widely available to consumers --- to allow BLV users to explore and learn about images.

Using touchscreen devices, researchers have created techniques that supplement alt text to offer BLV users a richer view of digital images~\cite{Morris2018, Kim2021, Oh2021}. Morris et al. looked at touchscreen-based image exploration as one of several techniques for providing BLV users with richer representations of images~\cite{Morris2018}. Systems, such as Lee et al.’s ImageExplorer~\cite{Lee2022a} and Microsoft’s Seeing AI~\cite{Microsoft2017}, use deep learning models to automatically detect important objects within an image and place bounding boxes around those objects. As users move their finger over the image and touch those bounding boxes, the app will verbally announce what those objects are. Touchscreen-based image exploration systems have also been studied as a technique for presenting maps and floor plans, both within physical world~\cite{Ducasse2018a, Poppinga2011a, Su2010} and video game~\cite{Nair2022} environments. 

Although not a complete replacement for physical tactile graphics, touchscreen-based image exploration systems present distinct benefits for BLV users, especially when compared with standard alt text and image descriptions. Prior work has found that these systems afford users a strong sense of the spatial layout and sizes of elements within the image~\cite{Lee2022a, Morris2018, Goncu2011}. Research has also found that allowing users to explore the image on their own terms via touch affords a sense of agency~\cite{Lee2022a}, and allows BLV users to more vividly imagine and understand what the image is showing~\cite{Ahmetovic2021}.

Prior work, however, has also found weaknesses in touchscreen-based image exploration tools. In particular, through their evaluations of ImageExplorer, Lee et al. found that exploring via touch imposed a high mental load on BLV users, causing users to express frustration while using the tool~\cite{Lee2022a}. Morris et al. additionally found that users were prone to missing information during exploration, and thus may receive incomplete information about the image~\cite{Morris2018}. Issues like these can make the experience of using touch to explore images more frustrating and time-consuming for users, though relatively little work has been done on thoroughly investigating and proposing solutions to these inefficiencies.

In this work, we consult with BLV users to deepen our understanding of the bottlenecks inherent in touchscreen-based image exploration systems, and then use our learnings to design and test tools that can potentially relieve these issues while preserving the benefits that touchscreen-based image exploration systems afford users.

%% file: sec03-formative.tex
We began by conducting a formative study to explore and identify major challenges faced by BLV users in the use of touchscreen-based image exploration tools. We created a smartphone-based image exploration system meant to faithfully recreate state-of-the-art approaches~\cite{Lee2022a, Microsoft2017, Morris2018, Ahmetovic2021}, and we had two totally blind users use it to explore several test images.

\subsection{Participants}

We recruited both participants (henceforth referred to as P1 \& P2) from a mailing list of BLV individuals who participated in our laboratory’s previous studies. P1 was aged 36-45, and P2 was aged 18-25. Both participants were male, described themselves to have “no usable vision,” and stated that their vision impairments developed at birth. P1 also reported minor hearing loss in one of their ears.

\subsection{Prototype Used}

We presented both participants with a smartphone-based image exploration system, based on similar state-of-the-art approaches in the literature such as Lee et al.’s ImageExplorer~\cite{Lee2022a}, Microsoft’s Seeing AI~\cite{Microsoft2017}, Morris et al.’s “spatial” interaction prototype~\cite{Morris2018}, and Ahmetovic et al.’s “hierarchical exploration” system~\cite{Ahmetovic2021}.

\begin{figure*}
    \begin{subfigure}{.32\textwidth}
      \centering
      \includegraphics[width=.98\linewidth]{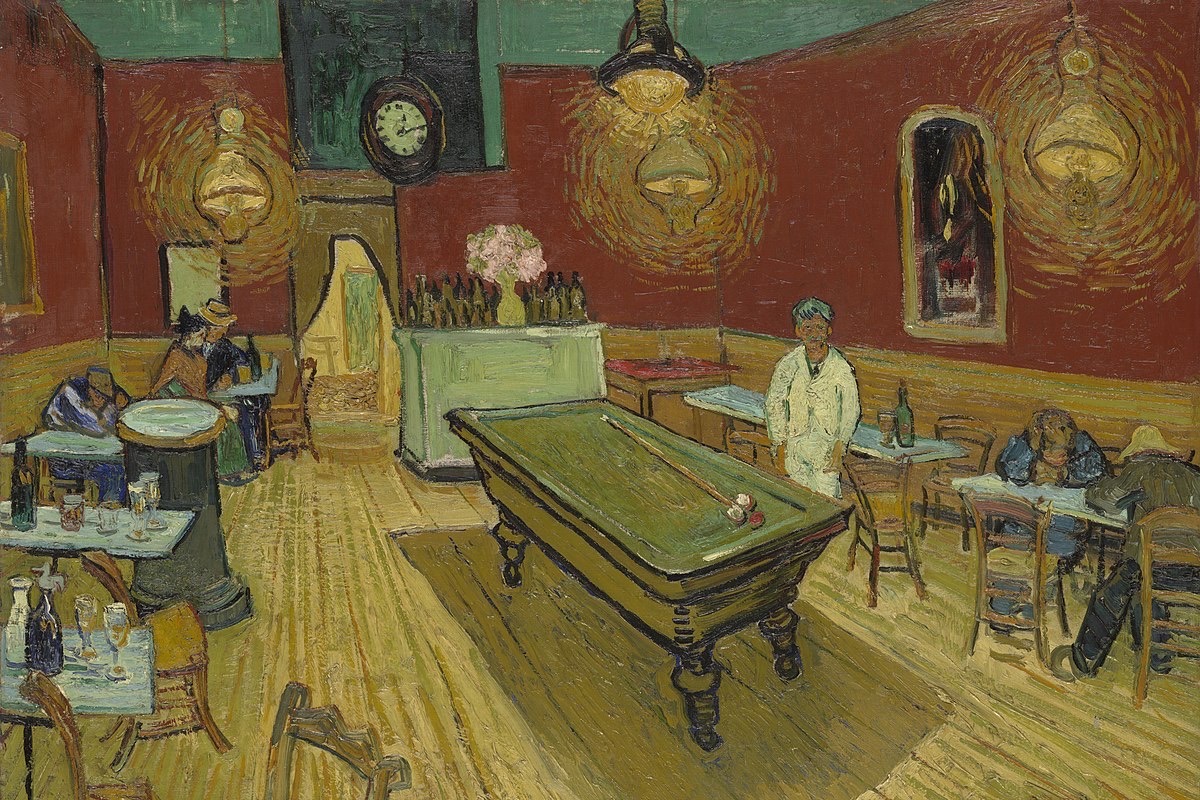}
      \caption{}
    \end{subfigure}%
    \begin{subfigure}{.32\textwidth}
      \centering
      \includegraphics[width=.98\linewidth]{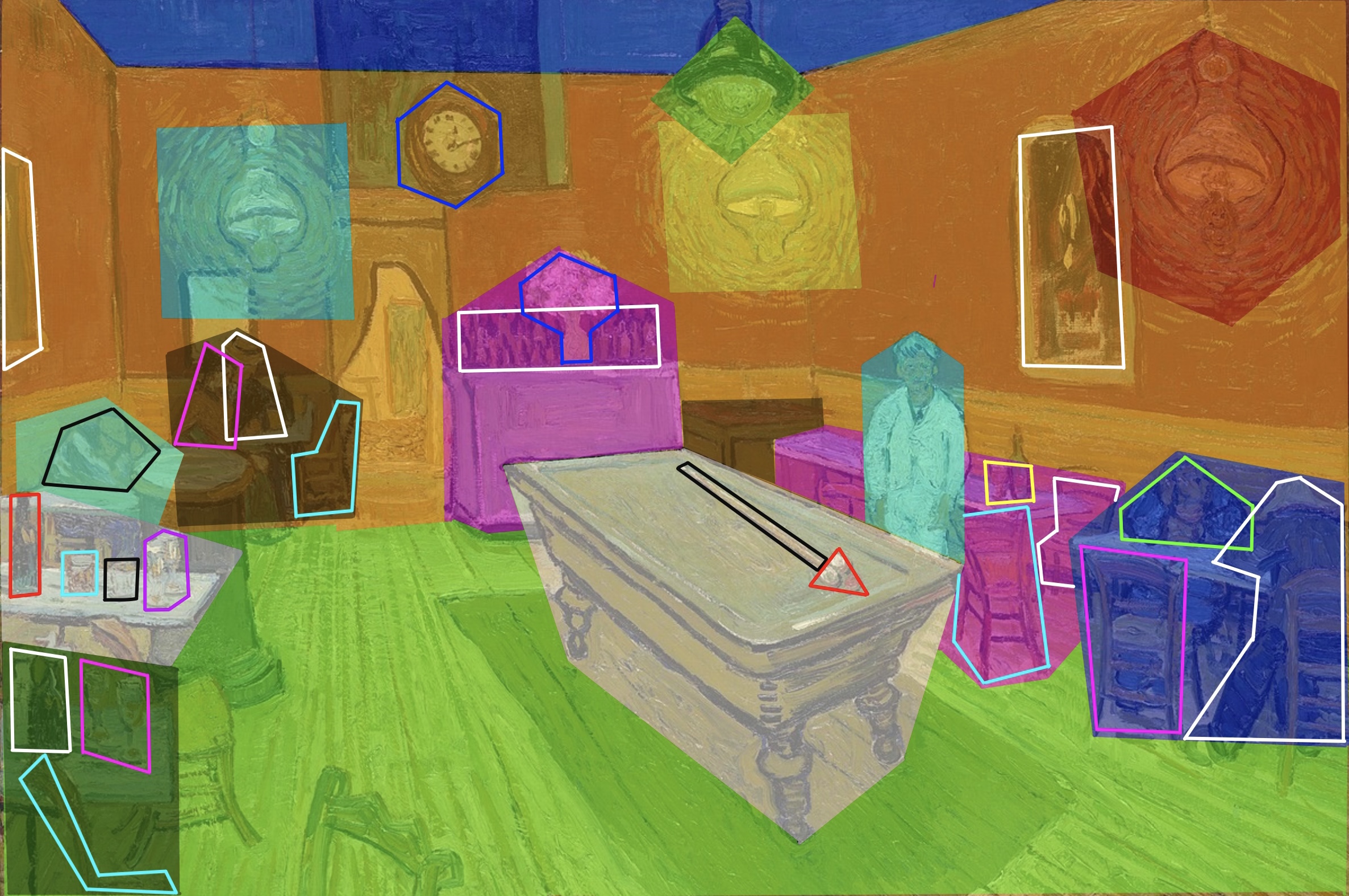}
      \caption{}
    \end{subfigure}%
    \begin{subfigure}{.32\textwidth}
      \centering
      \includegraphics[width=.98\linewidth]{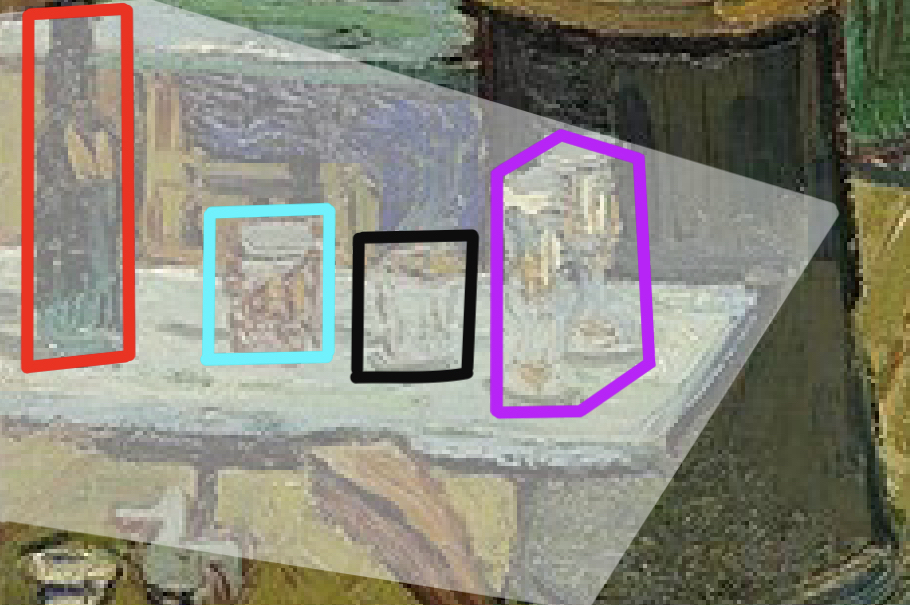}
      \caption{}
    \end{subfigure}
    \caption{\textit{The Night Café} by Vincent Van Gogh, one of the images we used in our formative study. (a) The original image. (b) Depiction of areas within \textit{The Night Café}. Colored planes represent areas on the top level. Outlines within those planes represent sub-areas within those top-level areas. (c) Zoomed in view of a table in \textit{The Night Café} (located on the far left side of the painting). Participants faced difficulties in surveying the small objects on the table.}
    \Description{Three panel image showing various views of a painting; the painting is of the interior of cafe with several tables, five people sitting at these various tables, and a pool table in the middle. The panels show the original painting, the painting with areas and sub-areas indicated, and a zoomed-in view of a specific table in the painting, which has several bottles and glasses on it.}
    \label{f:nightcafe}
\end{figure*}

In this system, images are divided into separate regions, or “areas,” which encompass different objects or components of the image. Users can survey the image by dragging their finger on the screen. When their finger enters an area, the system will speak the area's name (i.e., its label). Some areas additionally have “sub-areas;” the system will let users know that a top-level area has sub-areas by prompting --- “Double-tap to explore” --- after announcing the name of the area. Figure \ref{f:nightcafe}(b) shows an image divided into these areas and sub-areas.

Double-tapping allows users to “enter” the area, exposing its component sub-areas for surveying. The system plays a continuous warning tone whenever the user moves their finger off of the top-level area. Triple-tapping returns users to the top level of areas (i.e., the whole image). Upon triple-tapping, the system reminds the user of the number of unexplored areas remaining. This number decreases if users touch a top-level area with no sub-areas \textit{or} if they explore all sub-areas nested under a top-level area. If and when the user finishes exploring all areas and sub-areas, the system will announce that there are "no more unexplored areas."

We created the smartphone app using the Unity game engine~\cite{UnityTechnologies2020} which allowed us to create one app for both iOS and Android smartphones.


\subsection{Procedure}

We presented both participants with three test images presented in a random order. The three images were: 

\begin{itemize}
    \item a painting (\textit{The Night Caf\'{e}} by Vincent Van Gogh), 
    \item a photo (the cover of the album \textit{Abbey Road} by The Beatles), and 
    \item a floor plan of a generic condominium.
\end{itemize}

We picked these three images in order to represent a variety of image types. Before trying these three images, however, we had participants learn and test the tool within a trial image, which was another painting: \textit{Nighthawks} by Edward Hopper. We annotated the areas and sub-areas of these images by hand due to time constraints and to scope the work to focus solely on the technique of exploring images by touch, though prior work has also used AI-based methods~\cite{Microsoft2017, Lee2022a, Wu2017} and crowdsourcing~\cite{Su2012, Zhou2018} to annotate major areas within an image.

We focused on segmenting large objects/areas as top-level areas and attempted to have each area's bounds cover as much of the object(s) in question as possible. We determined containment primarily physically. However, we tended to fall back on the semantic relationships between objects if we thought that communicating such relationships would make understanding the image easier for the user. For example, in Figure \ref{f:nightcafe}, the people sitting \textit{around} a table as well as the objects \textit{on} the table are considered part of the table itself. Most objects attached to the wall are nested under the wall itself. However, the chandeliers are separate top-level areas since they are not, in fact, on the wall. The standing person is a separate top-level area: They are not interacting with the pool table or sitting at one of the tables.

We began the session with a pre-study questionnaire requesting demographic information. We then loaded the trial image onto the participant’s phone while explaining how the tool worked. After this trial image, we conducted a short interview with the participant about their prior experiences with exploring digital images. We then presented each test image one-by-one and allowed participants to explore the images at their own pace. Participants could end their exploration whenever they wanted. 

Before starting exploration, we presented participants with a brief description of the image so that they knew generally what the image was showing. We did this because prior work with similar systems has recommended providing a brief caption of the image before exploration in order to reduce the cognitive demands associated with gaining a global understanding of the image~\cite{Ahmetovic2021, Morris2018}; providing a brief caption beforehand thus removed any potential of this issue confounding our findings. After each image, we administered a questionnaire to participants to evaluate their experience using the tool on this image.

We performed this study completely remotely. We deployed the smartphone app to participants’ iOS and Android devices via the Google Firebase App Distribution service~\cite{FirebaseAppDist}. The app connected to a cloud backend which allowed us to control the participant’s app remotely.

To analyze sessions, we followed an inductive coding process that involved two members of the research team. Sessions were recorded with participants’ consent, and each coder went through transcripts and coded quotes and other events. Then, both coders iterated on the codes together until there was agreement that they could not iterate further.

\subsection{Challenges Faced by Users in Touchscreen-Based Image Exploration}

P1 and P2’s sentiments revealed three major challenges that BLV users face with touchscreen-based image exploration tools.

\subsubsection{Inability to locate regions.}\label{subsubsec:form-challenges-locate}

One of P1 and P2’s biggest complaints was that they missed out on exploring regions, either because they swiped over the region quickly and could not re-locate it or because they could not locate the remaining regions at all. This is an issue that has also been identified in prior work~\cite{Lee2022a, Morris2018}, and both P1 and P2 suggested that one way to fix this was to provide an overview of the image they are exploring:

\begin{quote}
    \textit{"I would have liked to have gotten an overview of where actually everything is, so that I knew I could look for the regions I missed. Because clearly, I missed a couple of details and don’t know what they were or where I could find them."} --- \textbf{P1}
\end{quote}

\subsubsection{Inability to survey efficiently.}\label{subsubsec:form-challenges-ineff}

Both participants commented that they were forced to survey the entire image inefficiently --- in this case, both participants kept swiping their finger back-and-forth across the entire image in a grid-like pattern. This was error-prone in that neither participant could keep their finger moving in a perfectly straight line across the screen, which caused them to lose track of where they were within the image. We found that both participants desired some form of direction --- possibly in the form of hints --- to direct them straight to regions of interest and thus make exploration much more efficient:

\begin{quote}
    \textit{"Instead of having to explore the whole image, I'd love if there was something that told you where to explore. Because how it is right now, you need to mechanically swipe over the whole [image], but there's no starting point and so you just need to randomly pick a point to start at, which can make the whole [exploration] process a bit annoying depending on where you begin."} --- \textbf{P1}
\end{quote}

\subsubsection{Difficulty surveying small and tightly clustered areas.}\label{subsubsec:form-challenges-smallareas}

The third major issue that both participants faced related to difficulties in surveying small regions on the image. These complaints were especially prevalent in \textit{The Night Caf\'{e}}, where the tables are designated as top-level areas and the objects on those tables are designated as sub-areas. Figure \ref{f:nightcafe} shows the structure of the regions in the painting. Both participants had difficulties in distinguishing between the sub-areas (i.e., the objects on the tables) and ascertaining their positions and shapes. These sub-areas were often clustered closely together and were very small:

\begin{quote}
    "\textit{Everything was, like, so bunched up here. I was almost trying to separate the elements and trying to get an image in my head of how things work, but it was difficult."} --- \textbf{P2}
\end{quote}

\subsection{Design Goals for ImageAssist}

The three challenges revealed in our formative study yield three major design goals that ImageAssist should address to make touchscreen-based image exploration much less frustrating:

\textbf{G1: Providing an overview of the image and the ability to identify and pinpoint areas of interest to the user.} Our formative study revealed that users had problems finding regions of interest to them or regions they have not yet swiped over at all. In order to help users find every region, ImageAssist should provide an accessible overview of the image that should also allow users to easily identify and pinpoint areas of interest to the user.

\textbf{G2: Giving users direction.} To help users survey images more efficiently, ImageAssist should provide users with a sense of direction, possibly by nudging users toward more prominent elements of the image so that users can survey those elements first.

\textbf{G3: Helping with small and tightly clustered areas.} In order to alleviate the difficulties that users face in finding small areas, ImageAssist should include a facility for making these small areas more accessible to users. 

%% file: sec04-probes.tex
In response to the results of our formative study, we created ImageAssist, a set of three tools intended to address the challenges that BLV users face in using touchscreen-based image exploration systems and, thus, enhance their experience using these tools. Figure~\ref{fig:teaser} depicts the three tools that comprise ImageAssist alongside the baseline touchscreen-based image exploration system we used in the formative study. They include a menu \& beacon tool, a hints tool, and a quadrant zoom tool.

In order to ensure the usability of each tool, we went through multiple design iterations with pilot tests to address the many open-ended design decisions that we had to make along the way. We conducted these pilot tests with two BLV people and two sighted-but-blindfolded people within our laboratory. (Note that our two BLV pilot testers were selected only for these pilot tests and did not participate in our formative or main studies.)

We intended for our tests with sighted-but-blindfolded participants to catch low-hanging fruit with respect to issues with the tools before piloting with our BLV participants. In the following subsections, we describe the design and implementation of the three tools in detail. We implemented ImageAssist within a smartphone app using the Unity game engine~\cite{UnityTechnologies2020}. 

\subsection{Menu \& Beacon Tool}
\label{sec:systems--mb}

\begin{figure*}
    \centering
    \begin{subfigure}{.45\textwidth}
      \centering
      \includegraphics[width=.98\linewidth]{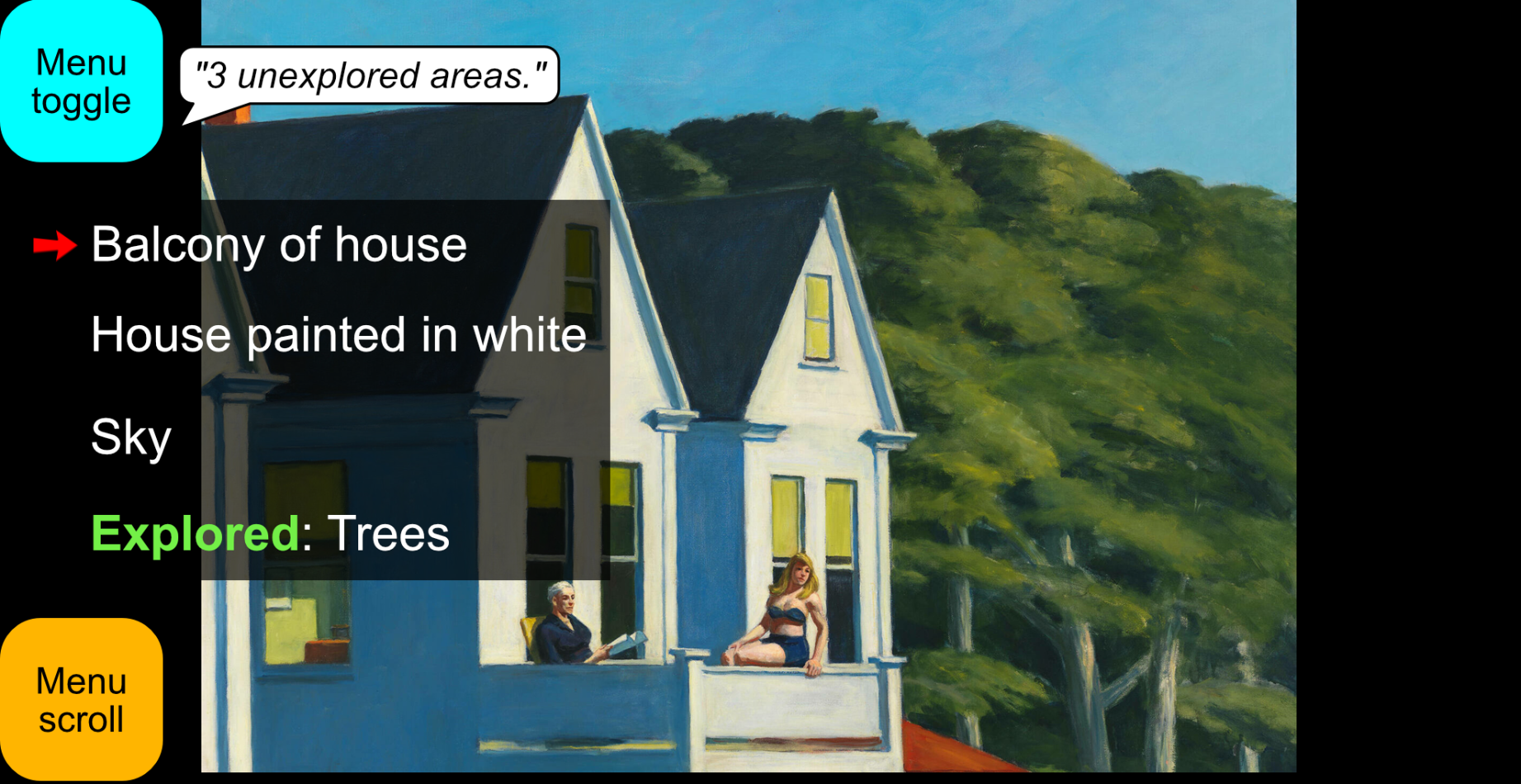}
      \caption{}
    \end{subfigure}%
    \begin{subfigure}{.45\textwidth}
      \centering
      \includegraphics[width=.98\linewidth]{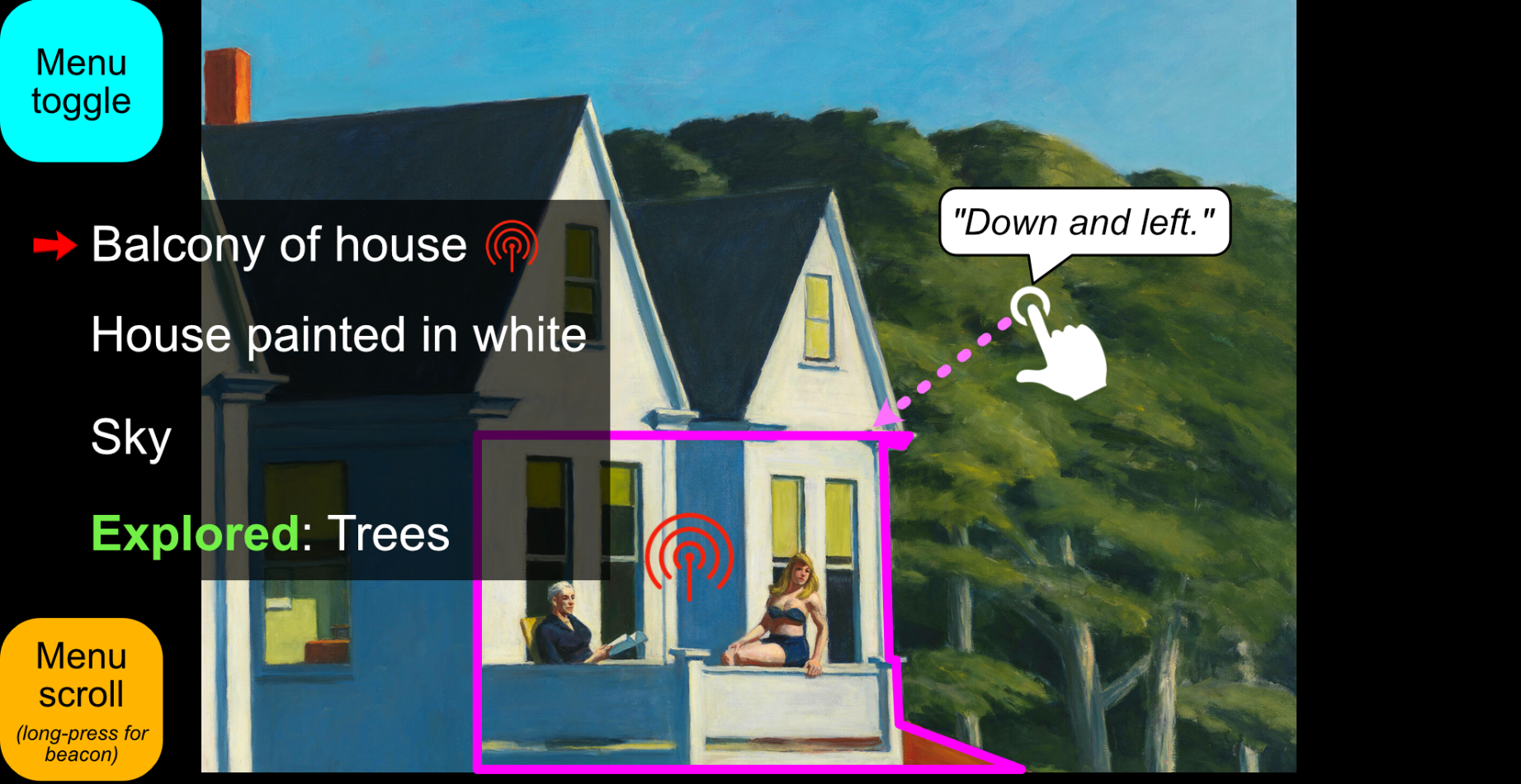}
      \caption{}
    \end{subfigure}
    \caption{ImageAssist's menu \& beacon tool. (a) Illustration of the menu. Users activate the menu by tapping the button on the upper-left portion of the screen, which also announces the number of unexplored areas. Users scroll through by tapping the button on the lower-left. The menu lists out all areas within the image, and areas that the user has already explored are moved to the end of the menu. (b) Illustration of the beacon. Users activate the beacon by holding the lower-left button after scrolling to an entry using the menu. The beacon uses a looping sound effect and announcements to direct the user to the target area.}
    \Description{Two panel image showing ImageAssist's menu and beacon system on a painting of a house alongside some trees under a blue sky. A menu showing specific areas in the image (such as, "balcony of house" and "house painted in white") appears in both panels.}
    \label{fig:menubeacon}
\end{figure*}

The menu \& beacon tool --- illustrated in Figure \ref{fig:menubeacon} --- announces all areas within an image using a simple audio menu, and directs the user to an area of their choice using an audio beacon. We designed and implemented this tool to achieve design goal \textbf{G1}: providing an overview of the image and the ability to identify and pinpoint areas of interest to the user. The menu \& beacon tool draws inspiration from research in navigation for BLV individuals in both physical world and video game environments~\cite{Wilson2007, MicrosoftResearch2018, Nair2021, Nair2022}.

The menu uses two buttons on the left-hand side of the user's smartphone screen to control an audio menu. When a user taps the button on the upper-left corner of the screen, the audio menu will open, and the app will announce the number of unexplored areas (or sub-areas if the user has already double-tapped into an area). The user can then tap the button at the lower-left corner of the screen to scroll down through the menu, which contains the names of areas within the currently active level.

At any point within the menu, users can hold the scroll button (i.e., the button at the lower-left of the screen) to activate the beacon tool. This action will set the target of an audio beacon to be the centroid of the currently selected area within the menu, and an announcement (“in beacon mode”) will play. The user then places their finger anywhere on the image, and the beacon will direct them to the target area. While the beacon is active, it provides two pieces of information to the user: 

\begin{enumerate}
    \item \textit{The direction in which the user’s finger should move to reach the target area.} This information is presented as a periodic directional announcement (e.g., “down and left”) --- as seen in Figure \ref{fig:menubeacon}(b).
    \item \textit{The distance from the user’s current touch point to the target area.} This information is presented by a series of beeps, whose frequency increases as the user moves their finger closer to the target area.
\end{enumerate}

When the user reaches any point within or on the boundary of the area, the beacon will automatically stop, and an announcement will play. The user can then continue exploring the image via touch as usual. The user can also manually cancel the beacon by holding the button at the lower-left corner of the screen again. 

In our initial design, the audio menu only announced the names of unexplored areas, and once an area was explored, it was removed from the menu. This was intended to promote more efficient exploration by decreasing the time spent scrolling. However, a pilot study participant mentioned that removing explored areas made it harder for them to remember the names of areas that they \textit{did} explore and that this prevented them from forming a more holistic mental picture of the image. Thus, in our final design, the names of explored areas are also displayed within the menu. To distinguish them from unexplored areas, they are prefixed with the word “explored” and are announced using a different voice in the menu.

\subsection{Hints Tool}
\label{subsec:tools--hints}

\begin{figure*}
    \centering
    \begin{subfigure}{.45\textwidth}
      \centering
      \includegraphics[width=.98\linewidth]{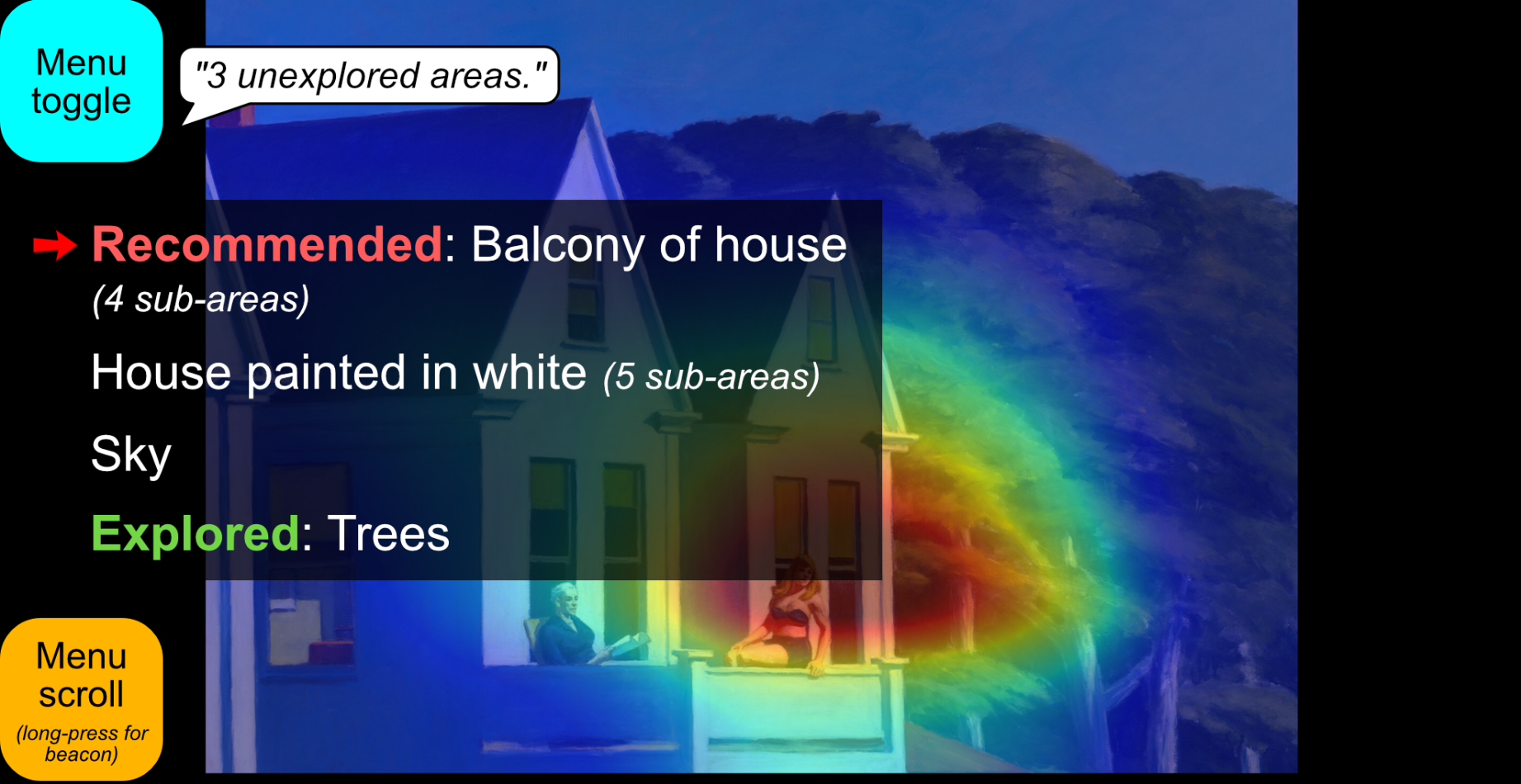}
      \caption{}
    \end{subfigure}%
    \begin{subfigure}{.45\textwidth}
      \centering
      \includegraphics[width=.98\linewidth]{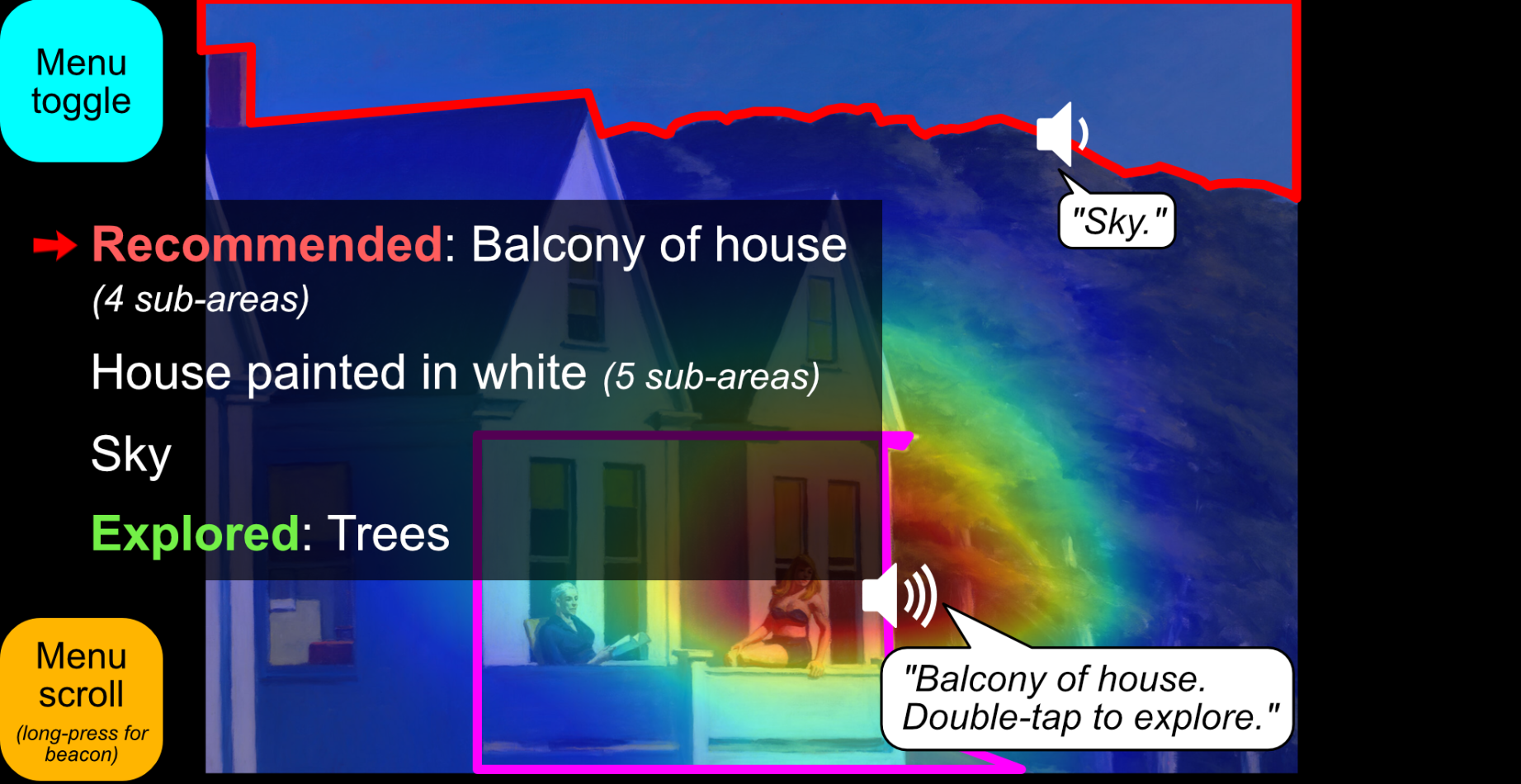}
      \caption{}
    \end{subfigure}
    \caption{ImageAssist's hints tool. (a) Illustration of the enhanced menu hint, which adds recommended areas to the beginning of the menu and appends the number of sub-areas to each area's entry within the menu. (b) Illustration of the prominence indicator hint. The class activation map for the image is used to determine the prominence of each area within the image; the more prominent the area, the louder its name will be announced when touching it.}
    \Description{Two panel image showing ImageAssist's hints system on a painting of a house alongside some trees under a blue sky. In both panels, the painting is overlaid by a heatmap whose brightest area lies on the balcony of the house.}
    \label{fig:hint}
\end{figure*}

The hints tool --- shown in Figure \ref{fig:hint} --- conveys three types of hints: an enhanced menu, a prominence indicator, and a first touch indicator. We designed and implemented this tool to achieve design goal \textbf{G2}: giving users direction. 

\begin{figure*}
    \centering
    \begin{subfigure}{.45\textwidth}
      \centering
      \includegraphics[width=.98\linewidth]{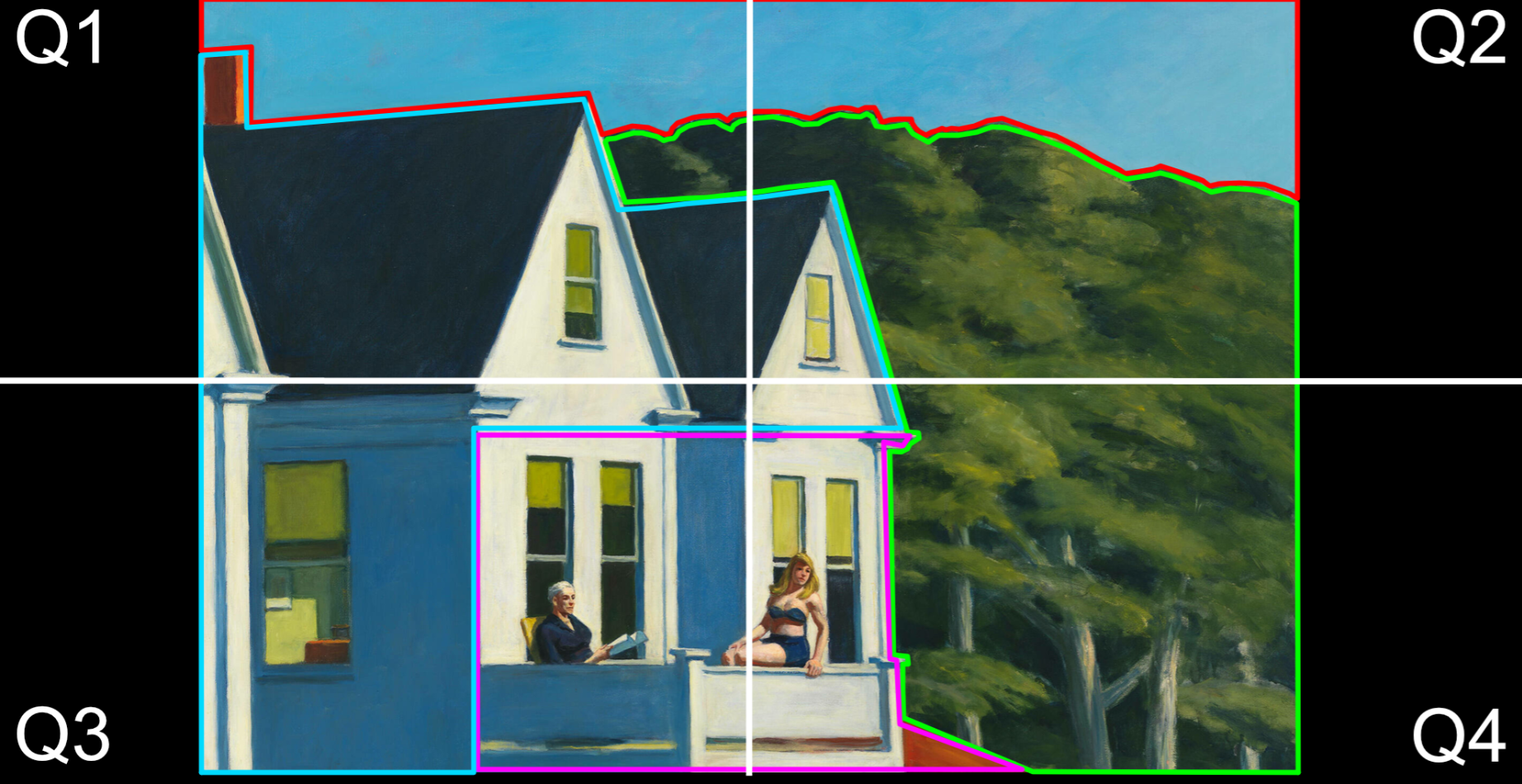}
      \caption{}
    \end{subfigure}%
    \begin{subfigure}{.45\textwidth}
      \centering
      \includegraphics[width=.98\linewidth]{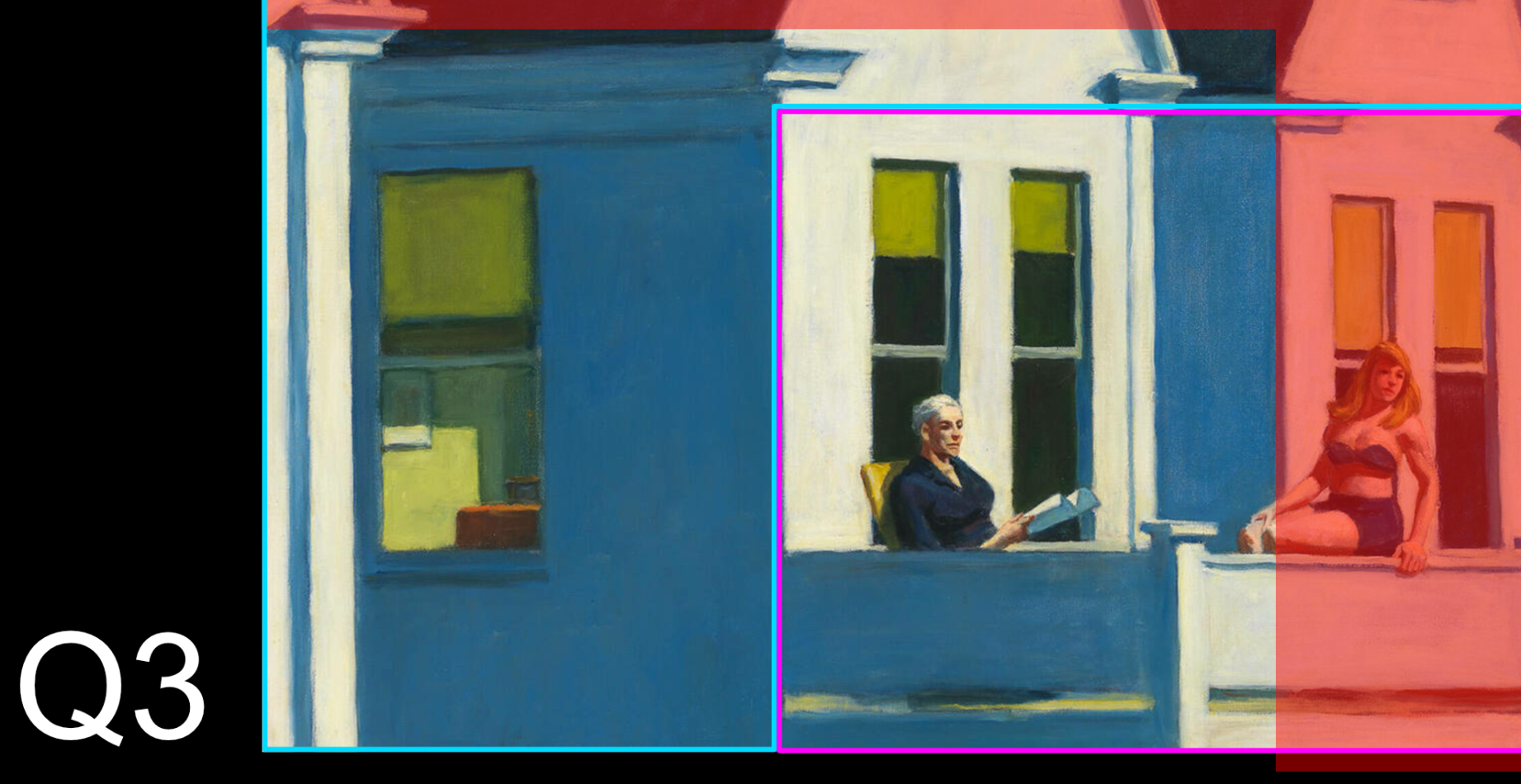}
      \caption{}
    \end{subfigure}
    \caption{ImageAssist's quadrant zoom tool. (a) The image is broken up into four quadrants (labeled Q1 to Q4 in the figure). (b) The user's view after zooming into the bottom-left quadrant (Q3) of the image. They can use their finger to survey just that quadrant. Parts of the image shaded in red bleed into other quadrants, and the user cannot survey them until they zoom back out --- the tool will play an announcement and a warning tone if the user's finger enters those red-shaded areas.}
    \Description{Two panel image showing ImageAssist's quadrant zoom system on a painting of a house alongside some trees under a blue sky.}
    \label{fig:zoom}
\end{figure*}

The \textit{enhanced menu} preserves the menu and beacon functionalities as described in Section \ref{sec:systems--mb}. However, this new tool modifies the list of areas as follows:

\begin{enumerate}
    \item A list of areas that users are recommended to explore appear at the beginning of the menu. The names of these areas are prefixed with the word “recommended” as users scroll through the menu. We labeled an area as "recommended" if the research team unanimously agreed that the area is a visual focus of the image, and that exploring this area is critical for understanding the image --- for example, if the element is a person who is a focus of the given image, then they will be "recommended."
    \item For areas that have sub-areas, the number of sub-areas is announced after the area's name. With the exception of the recommended areas, all areas within the image are sorted in descending order by the number of sub-areas they contain, with ties broken alphabetically. Recommended areas themselves are sorted alphabetically.
\end{enumerate}

These changes are designed to highlight the important parts of an image so that users can choose to explore just those areas to gain a quick understanding of the image. That is, the recommended areas provide users with the \textit{semantic} prominence of areas within the image. For example, in a floor plan --- such as the one in Figure \ref{fig:main-study-images}(d) --- it is often easier for a user to understand the floor plan if they know where the entrance of the area is. They can then survey the areas within the floor plan relative to the area's entrance. As such, the enhanced menu will surface that area first.

The \textit{prominence indicator} changes the volume of spoken area names to correspond to how prominent the area is within the image. To determine the prominence of areas, we used Smooth Grad-CAM++~\cite{Fernandez2021, Omeiza2019} to generate a class activation map (CAM) for each image. The heatmap shown in Figure \ref{fig:hint} is an example of a CAM, which visualizes the activation of the underlying network through which the image was run. We ran images through a residual neural network (ResNet-18)~\cite{He2015, PyTorchTeam2017} that was pre-trained on the ImageNet dataset~\cite{Deng2009}; residual neural networks and ImageNet have been used in prior work exploring prominence in the context of saliency and eye fixation on an image~\cite{Chen2020, Figurnov2017, Kummerer2015, Lv2019}.

A CAM provides an activation value for each image pixel; we compute each area's prominence by averaging its component pixels' activation values. Prior work has found that there may be discrepancies in area recognition between human eyes and neural networks~\cite{vanDyck2021}; however, the research team agreed with all CAM outputs for the study images.

The prominence indicator provides users with the \textit{visual} prominence of areas within the image. For example, the heatmap in Figure \ref{fig:hint} shows that the balcony is more prominent than the sky, because the balcony and the people on it are the visual focus of the image. Therefore, when a user touches both areas, ImageAssist will announce "Sky" at a much lower volume than "Balcony." 


The \textit{first touch indicator} is a small sound effect that alerts the user when it is their first time touching an area. The sound effect is played just before the area's name is spoken. This indicator helps the user keep track of which areas they have already touched. While users can also scroll through the menu to hear the names of explored and unexplored areas, the first touch indicator provides a shorter and more straightforward alternative.

\subsection{Quadrant Zoom Tool}
\label{subsec:tools-zoom}

The third tool we created is the quadrant zoom tool --- illustrated in Figure \ref{fig:zoom} --- which allows users to zoom into a portion of the image. We created this tool to address design goal \textbf{G3}: helping with small and tightly clustered areas.

As Figure \ref{fig:zoom} illustrates, our quadrant zoom tool divides the image into four equally sized quadrants that the user can zoom in and out from. Users double-tap any quadrant using two fingers to zoom into that particular quadrant, and triple-tap with two fingers anywhere on the screen to zoom back out. An announcement confirms the zoom action. While zoomed into a quadrant, users can use their finger to survey as usual, including double-tapping into areas that have sub-areas. It is important to note that an area may only partially lie within a single quadrant, as is the case with the balcony in Figure \ref{fig:zoom}(b). When zoomed in, the user will only be able to ``see'' what lies in the currently zoomed-in quadrant and will need to zoom out and zoom back in to the other quadrant if they want to explore the entire balcony up close.

The quadrant zoom tool represents our second attempt at designing non-visual zooming functionality. Our initial design, which we abandoned after pilot testing with users, was a combination zooming + panning tool that operated similarly to how zooming and panning work for sighted users on touchscreens. In the initial tool, users could perform a pinch gesture with two fingers to zoom in and out of the image, and could slide along the top/bottom and left/right sides of the screen to pan horizontally and vertically, respectively. Extensive announcements about the current viewport window size accompanied these operations.

Our pilot tests showed that users were confused by the concept of pinching-to-zoom and panning across an image since they had no frame-of-reference for the zoom levels and pan positions that were being announced. It was also cognitively demanding for them to keep track of these values via the announcements. Although prior work has investigated other methods for non-visual zooming and panning~\cite{Palani2016, Palani2017}, we ended up settling on our quadrant-based technique to keep the tool simple for our users.

%% file: sec05-us-overview.tex
Using the three tools we created, we performed a user study to investigate how well and in what ways they enhance touchscreen-based image exploration systems for BLV users. Our user study was fully remote and involved nine BLV participants. In the following subsections, we describe our study procedure.

\subsection{Participants}
\label{subsec:us-ptcpts}

Our nine BLV participants included the two participants from the formative study (referred to, again, as P1 \& P2 in this study) whom we invited back to test our new tools. We recruited the other seven from our laboratory's mailing list and from social media sites. Table \ref{tab:demographics} shows demographic information for all nine participants. All participants reported themselves as being very experienced with the smartphone they were using for the study (4+ on a 5-point Likert scale). 


\subsection{Procedure}

We began the study session by administering a pre-study questionnaire requesting demographic information. After the pre-study questionnaire, we began the two-part study, which consisted of a tutorial phase and a testing phase.

\begin{table*}[t]
\begin{tabular}{c|c|c|l|l|l}
\textbf{ID} & \textbf{Gender} & \textbf{Age}   & \textbf{Vision Impairment (VI)}                    & \textbf{VI Onset} & \textbf{Hearing Impairment}       \\ \hline \hline
P1 & M      & 36-45 & Total blindness                           & Birth    & Slight loss in right ear \\
P2 & M      & 18-25 & Total blindness                           & Birth    & None                     \\
P3 & M      & 26-35 & Total blindness                           & Birth    & None                     \\
P4 & M      & 26-35 & Total blindness                           & Birth    & None                     \\
P5 & M      & 26-35 & Total blindness                           & Birth    & Slight loss in left ear  \\
P6 & F      & 26-35 & Total blindness                           & Birth    & None                     \\
P7 & M      & 18-25 & Total blindness                           & Birth    & None                     \\
P8 & M      & 26-35 & Total blindness                           & Birth    & None                     \\
P9 & M      & 18-25 & Blind in one eye; reduced acuity in other & Birth    & None                    
\end{tabular}
\vspace{2mm}
\caption{Demographic information of all nine main study participants, obtained during pre-study.}
\label{tab:demographics}
\end{table*}

During the tutorial phase, participants learned how to use each tool within a single trial image (here, a simple painting: \textit{Second Story Sunlight} by Edward Hopper). For each tool, the study facilitator read out a description of the tool and instructions on using it. Afterwards, we allowed participants to use the tool for as long as they wanted on the trial image before moving on to the next tool.

Participants learned how to use four tools: the baseline tool, which is the same touch-based image exploration system we used in the formative study (i.e., the state of the art); the menu \& beacon tool; the hints tool; and the quadrant zoom tool. Note that, in the quadrant zoom condition, we gave users access to the menu \& beacon tool as well as the hints tool. However, participants made very little use of these extra tools during the quadrant zoom condition, and as such, we present results for the quadrant zoom condition as if we had tested the tool independently.

After the tutorial session, we moved on to the testing phase. Each participant tested with the following four images: 

\begin{itemize}
    \item a painting (\textit{The Art of Painting} by Johannes Vermeer),
    \item a stock photo (a scene of a summer day on an urban greenway),
    \item a historical photo (a photo taken in the 1930s of a man holding up a newspaper), and
    \item a floor plan of a generic condominium --- different from the one we used in the formative study.
\end{itemize}

\begin{figure}[t]
    \centering

    \begin{subfigure}{0.23\textwidth}
        \centering
        \includegraphics[width=\textwidth]{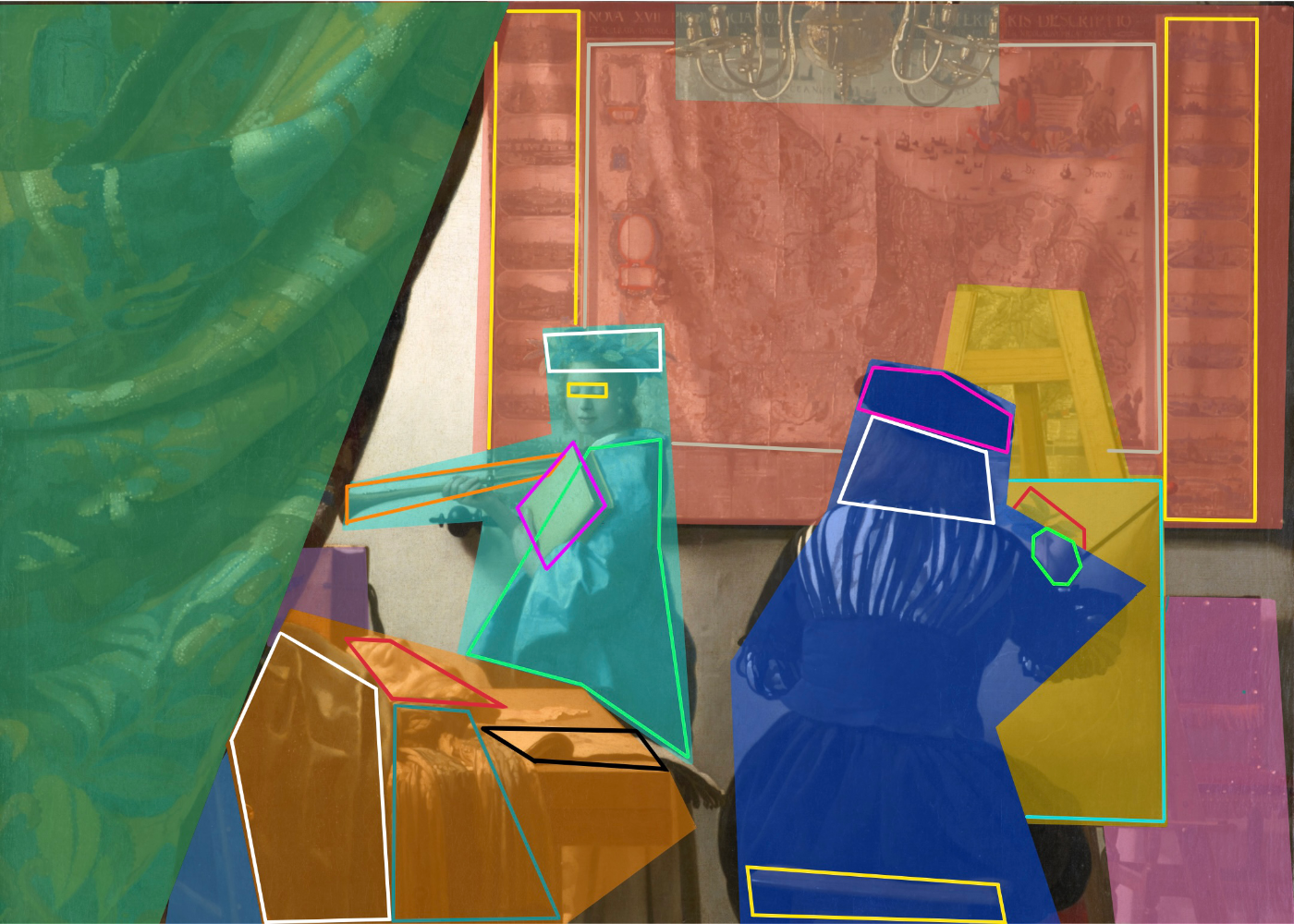}
        \caption[]{Painting}    
    \end{subfigure}
    \hfill
    \begin{subfigure}{0.23\textwidth}  
        \centering 
        \includegraphics[width=\textwidth]{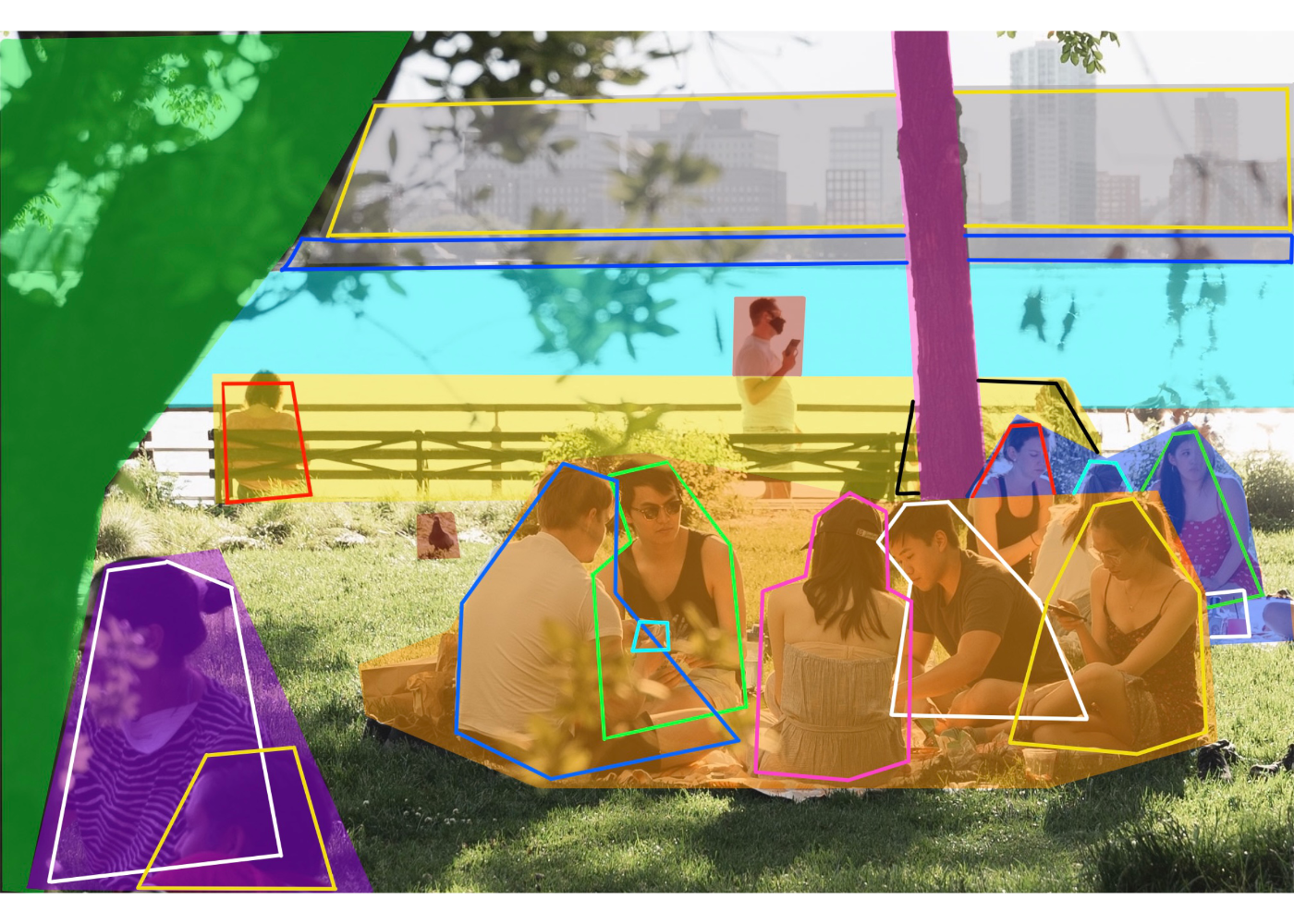}
        \caption[]{Stock photo}  
    \end{subfigure}
    
    \vskip\baselineskip
    
    \begin{subfigure}{0.23\textwidth}   
        \centering 
        \includegraphics[width=\textwidth]{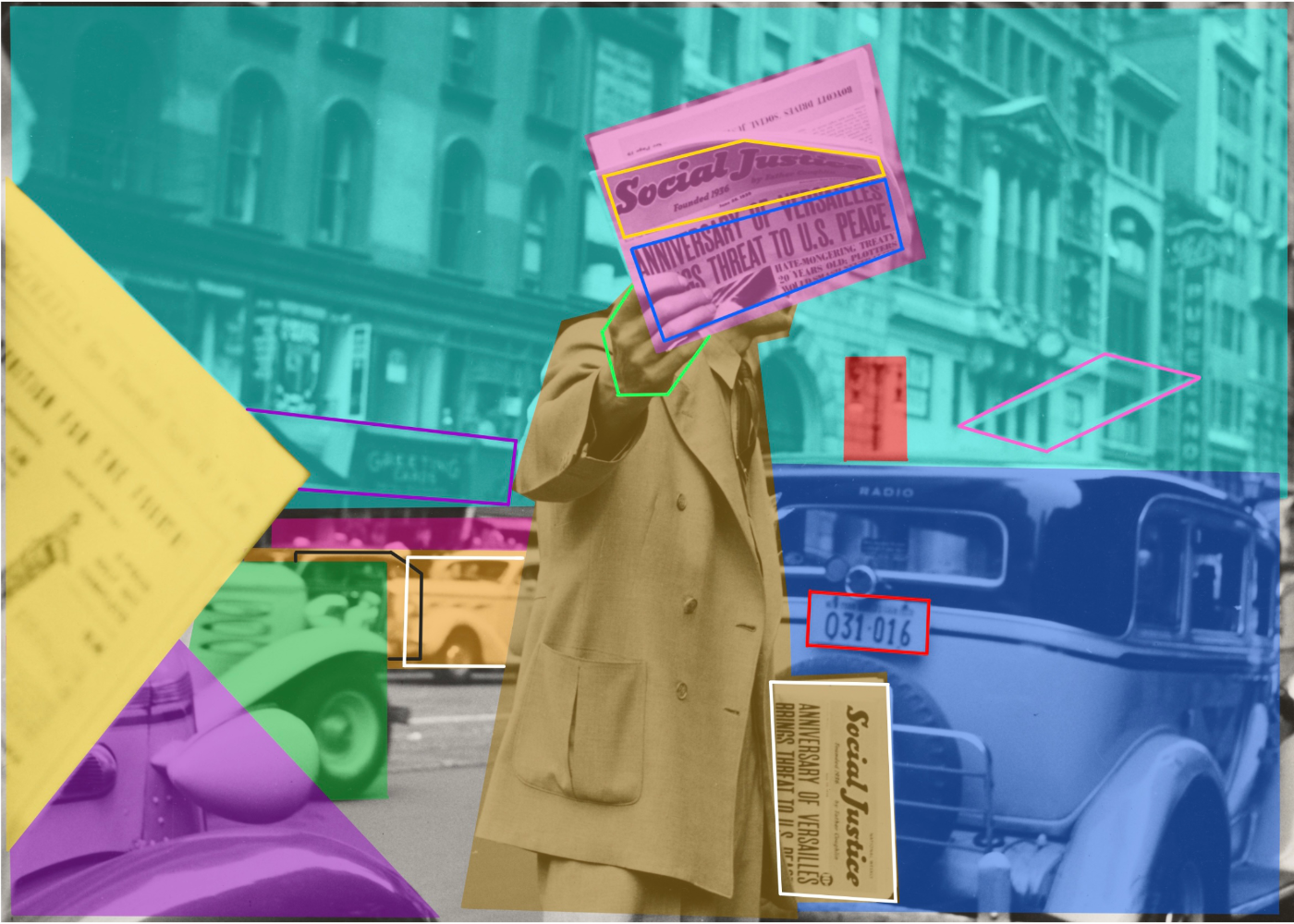}
        \caption[]{Historical photo}   
    \end{subfigure}
    \hfill
    \begin{subfigure}{0.23\textwidth}   
        \centering 
        \includegraphics[width=\textwidth]{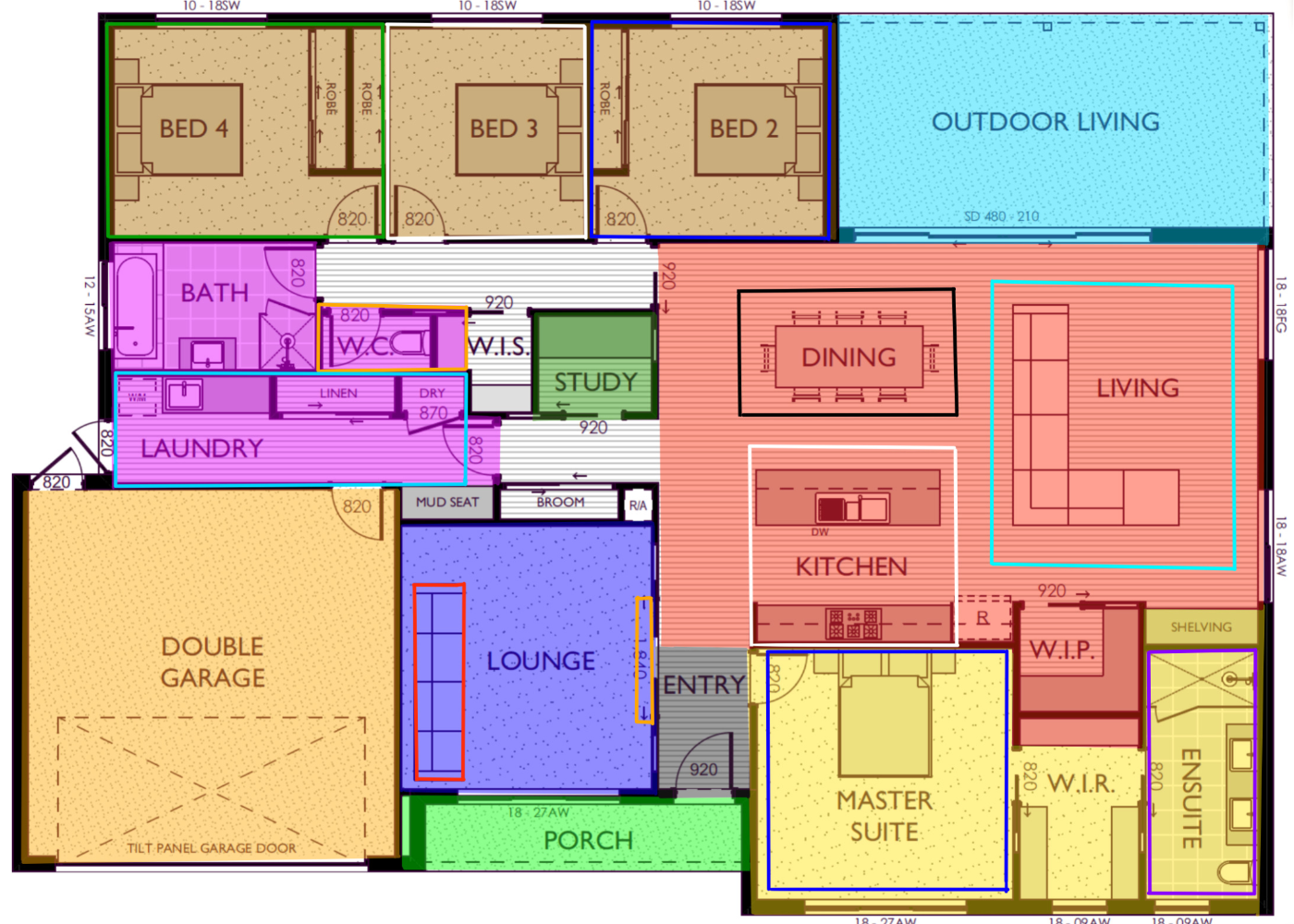}
        \caption[]{Floor plan}    
    \end{subfigure}
    
    \caption{The four test images we used in our user study. Solid-colored planes represent areas on the top level, while polygonal outlines within those planes represent sub-areas. (a) A painting (\textit{The Art of Painting} by Johannes Vermeer). (b) A stock photo (a scene of a summer day on an urban greenway). (c) A historical photo (a photo taken in the 1930s of a man holding up a newspaper). (d) A floor plan of a generic condominium.}
    \Description{Four panel image showing the four images we used in our main user study with areas and sub-areas indicated. The painting is of a man painting a woman in a medieval setting. The stock photo is of a group of friends having a picnic together alongside the Hudson River; there are many other people in this park as well. The historical photo is a black-and-white photo of a man holding up a newspaper in front of his face. The floor plan is of a condominium with many rooms.}
    \label{fig:main-study-images}
\end{figure}

Figure~\ref{fig:main-study-images} shows the four images we used with the areas and sub-areas on those images colored in. As with the formative study, we chose these four image categories to see how well ImageAssist works with a diverse set of images and to see what value it adds for BLV users across these types of images. Similar to the formative study, we annotated the areas and sub-areas manually.

Participants used each tool with one of the four images listed above. In order to control for order effects, we counterbalanced the order of the tools as well as the order of the images using Latin squares in order to see how each tool performs across images and how participants react to unexpected changes in the tools they are provided.

For each (tool, image) combination, we began by presenting participants with a brief caption so that they knew generally what the image was showing --- similar to what we did in the formative study. Users \textit{then} surveyed the image using the tool for as long as they desired, after which we administered a post-image questionnaire. The questionnaire asked participants to describe the image they had just surveyed and to relay what they thought was the most important part of the image. We then gauged participants’ subjective impressions of the tool used via open-ended responses as well as 20-point Likert scales similar to those found on a NASA TLX form~\cite{Hart1988}. At the very end of the study, we administered a post-study questionnaire gauging participants’ final impressions of the tools.

The technical setup for this main study was identical to that of the formative study, where we distributed the app to participants’ iOS and Android devices and had the app connect to a cloud backend, allowing us to control it remotely. Due to quirks in how Unity handles touch events, participants had to enable VoiceOver's "Direct Touch mode" on iOS or had to temporarily disable TalkBack on Android in order to use the app. However, using the phone's native screen reader was not necessary for this app — our placement of the menu's buttons at the corners of the screen (alongside the fact that these buttons emitted audio feedback when tapped) meant that participants could easily find them, and any app state changes were handled by us remotely.

To ensure a level playing field, we blanked out the app’s screen for all participants, so that users would not be able to see the images at all even if they possessed some level of vision. This, in particular, affected P9, who possessed residual vision in one of their eyes. P9 started the study by saying they would not use these tools much in real life because they had some residual vision. Later in the session, however, they acknowledged that their visual acuity was very low and that they could see themselves using some of these tools.

\subsection{Data Collection and Analysis}

We administered all questionnaires by having the facilitator read out each question and input the participant’s response into an internal Google Form. We recorded all sessions with participants’ permission for transcription purposes. We also obtained raw data of participants’ actions using the tools by capturing logs via our cloud backend. 

As in the formative study, we followed an inductive coding process to identify themes in participants' open-ended sentiments. Three members of the research team went through session transcripts and coded quotes and other events. Then, all members iterated on the codes together until there was agreement that they could not iterate further.

%% file: sec06-us-findings.tex
Here, we report the findings from our user study. The following subsections dive into our findings. Subsections \ref{subsec:results-mb}, \ref{subsec:results-hints}, and \ref{subsec:results-zoom} go into findings specific to each of ImageAssist's component tools. Subsection \ref{subsec:results-general} presents some additional findings that compare the ImageAssist suite with the baseline.

\subsection{Menu \& Beacon Tool}
\label{subsec:results-mb}

\begin{figure}
    \centering
    \includegraphics[width=0.45\textwidth]{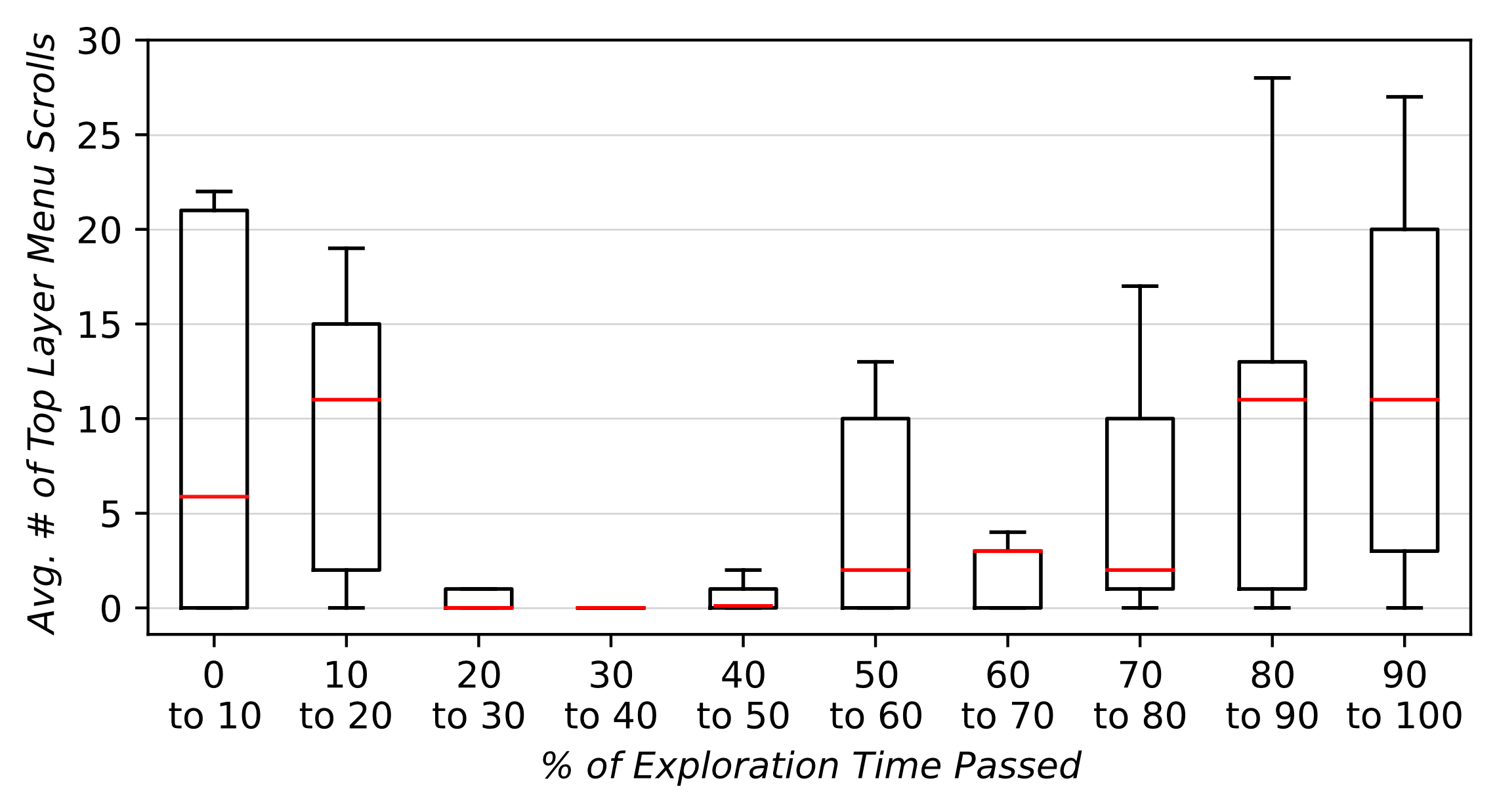}
    \caption{Box plots showing average number of times participants pressed the menu scroll button while they were on the top level of the image at various points of their image exploration sequences. Red lines indicate medians. These values are from the menu \& beacon condition only. Participants tended to scroll through the menu more at the beginning and end of the exploration process.}
    \Description{A set of box plots whose y-axis is representing the average number of times participants tapped the menu scroll button and whose x-axis represents different points of the exploration process --- in terms of percentage of the total time spent exploring an image during this study. The number of menu scrolls are high within the first 20\% of the total exploration time, goes down, and peaks again within the final 20\% of the total exploration time.}
    \label{fig:menuscrolls}
\end{figure}

Our study revealed insights about how the menu \& beacon addresses our first design goal: providing an overview of the image and the ability to pinpoint areas of interest. Figure \ref{fig:menuscrolls} shows the average number of times participants pressed the menu scroll button while on the top level throughout their time exploring the image. From the figure, we can see that participants tended to scroll through the menu more at the beginning and the end of the exploration process. This observation, as well as participants’ feedback on the menu, sheds light on how the menu served as a ``directory'' that allowed participants to obtain information about the image without needing to touch it. P5 relayed the following sentiment:


\begin{quote}
    \textit{“[The menu] shows you where things are without you actually having to explore it, which is good if you want a quick overview and you just want to know what is all there.”} --- \textbf{P5}
\end{quote}

Despite that sentiment and our instructions to participants that they could stop exploring the image whenever they desired, participants tended to touch \textit{more} areas within images when using the menu \& beacon tool than when they used the baseline tool by itself. In terms of the number of areas they touched, participants --- on average --- touched $84 \pm 12\%$ (median: 88\%) of areas on the image when using the baseline tool. This is in contrast to $94 \pm 8\%$ (median: 100\%) of areas touched when using the menu \& beacon tool.

This result suggests that most users do not treat the menu as an alternative to touch exploration, but rather as a tool that supplements touch exploration. This also suggests that the presence and use of the menu does not preclude users from exploring the image via touch.


Although most users chose touch as their primary method of exploration, some participants such as P7 noted that the menu is especially helpful at the beginning of exploration, when they have little information to go off of: 

\begin{quote}
    \textit{“Having the menu allows me to iteratively go through it and know that there are N number of areas or items that I need to go look for, so I'm not on a wild goose chase.”} --- \textbf{P7}
\end{quote}


For some participants, exploration became much harder without the menu keeping track of areas that were explored and unexplored. This was especially the case for five participants, who tested the menu \& beacon tool before the baseline tool as part of our counterbalancing scheme. All five of them expressed their frustration when the menu was removed. P1 even pointed out that they lost interest in exploring the image in detail when they could not find the areas they missed when using the baseline tool on the painting:

\begin{quote}
    \textit{“There was a point where I skipped a couple of details on the woman because I just couldn't find them. And I was just like, 'I don't even care, I just want to get through the picture.'”} --- \textbf{P1}
\end{quote}

Some participants mentioned that having a list of explored areas within the menu helped strengthen their understanding of the image. P6 said that they asked themselves reflective questions when scrolling through the menu to see if they understood the image, and P2 stated the following:

\begin{quote}
    \textit{“I really like that [the menu] keeps track of what you already explored. It is a very nice way to help you remember everything.”} --- \textbf{P2}
\end{quote}

Participants' sentiments above are corroborated by the large increase of menu scrolls at the end of exploration, as shown in Figure \ref{fig:menuscrolls}. At this point, by definition, users have explored most of the areas that they could find and are likely looking for new areas to complete their mental visualization of the image. The increase in menu scrolls suggests that repeatedly scrolling through the menu --- and over the areas they have and have not yet explored --- helped users solidify their understanding of the image.

The beacon helped further reinforce participants' understanding of the image, by enabling participants to find specific areas-of-interest in a precise manner. Participants placed, on average, 2.9 beacons during every round that featured a menu \& beacon tool. Given that there are many more regions within every image, this indicates that participants did not solely rely on the beacon to look for every area within the image and that it served as a supplement to touch-based exploration (similar to the menu). P9 summarized participants' sentiments on the beacon --- citing that the beacon was necessary in finding areas they may have missed:

\begin{quote}
    \textit{``The beacon being one of the primary ways to figure out what's going on means that you have a decent --- you have a reasonable shot of finding stuff on the image.''} --- \textbf{P9}
\end{quote}

\subsection{Hints Tool}
\label{subsec:results-hints}

Five participants stated that the first touch indicator was helpful in quickly reminding them of what areas they had and had not seen already without needing to scroll through the menu. These indicators decreased users’ mental load because they did not have to remember whether or not they had already seen the area and did not have to scroll through the menu to remind themselves. For example, P1, as part of our counterbalancing scheme, used the hints tool in the first round and then used the menu \& beacon tool (i.e., without the hints) in the second round. They stated the following after using the menu \& beacon tool in the second round on the stock photo image:

\begin{quote}
    \textit{“I missed having the hints. I found those to be very helpful, especially in a very busy picture like this one. it would have been nice to have the pings to quickly let me know that I found everything, which would have made things a bit easier.”} --- \textbf{P1}
\end{quote}

P1’s sentiment confirms that adding a simple indicator for newly discovered areas while the user is touching the screen can help immensely in decreasing their mental load.

Participants generally thought that the prominence indicator prompted them to think about the image differently --- allowing them to form their own interpretation of the image. One such participant, P7, relayed the following sentiment:

\begin{quote}
    \textit{“Unfortunately, as a blind person, I'm sort of confined to the beauty of the natural world around me to how others perceive it, which is something I think a lot of [BLV] folks, you know, struggle with. [...] And so having a tool like [the prominence indicator] is actually pretty neat, because it gives you the insight to understand what other people might be seeing, while also telling you how things are laid out so you can see things for yourself.”} --- \textbf{P7}
\end{quote}

Participants also believed that having the recommended areas in the menu was helpful in drawing their focus directly to those particular areas. With these explicit recommendations, users did not have to expend energy figuring out and finding the most important areas via touch. Some participants also remarked that both prominence and recommendations were pieces of information that they would otherwise not have known.

\begin{figure}
    \centering
    \includegraphics[width = 0.45\textwidth]{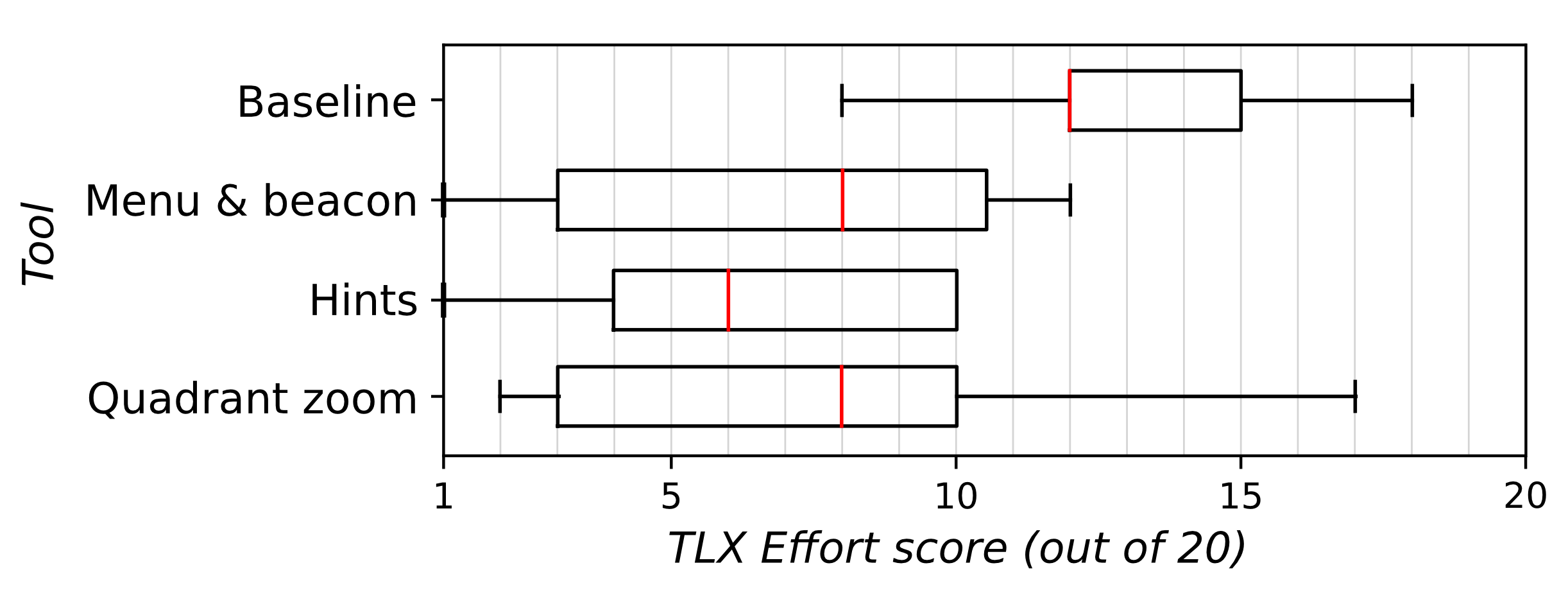}
    \caption{Box plots showing TLX effort scores for all four tools we tested. Red bars indicate median scores. The hints tool received the lowest median effort scores, indicating that the hints tool that did not impose a higher mental load on users.}
    \Description{Box plot of TLX effort scores on a scale of 1 to 20. The baseline tool has the highest median effort scores (with 12) over the quadrant zoom tool (8), menu and beacon (8), and hints (6).}
    \label{fig:tlxefforthint}
\end{figure}

Although the hints provided direction, we found that they did not impose a higher mental load on users. Figure \ref{fig:tlxefforthint} shows a box plot of TLX effort scores provided by participants for all four tools we tested. We can see that the hints tool received the lowest median effort scores out of all four tools.

We also noticed an interesting trend in the post-image questionnaire when we asked participants to tell us what they thought the most important part of the image was and why they gave that answer. We initially believed that, after using the hints tool, participants would answer this question with an area marked as recommended or prominent by the hints tool and cite the tool as why they picked that area. However, we observed that, while many participants would answer the question with the area(s) recommended by the enhanced menu and prominence indicators, they would not cite the hints as the reason. For example, P9 gave the following answer after using the hints tool on the painting:

\begin{quote}
    \textit{“I'm guessing the artist is the 'correct' answer because I think the menu recommended that I visit the ‘artist with a paintbrush.’ But I think the woman [is very important] because I feel like there was a large area dedicated to her.”} --- \textbf{P9}
\end{quote}

One explanation for this is that participants could have used the hints as confirmation for their own initial guesses --- that is, participants may have initially deduced the most important parts of the images from characteristics such as area size, area position, and number of sub-areas, and then used the hints to confirm their guesses.

\subsection{Quadrant Zoom Tool}
\label{subsec:results-zoom}

\begin{figure}
    \centering
    \includegraphics[width = 0.45\textwidth]{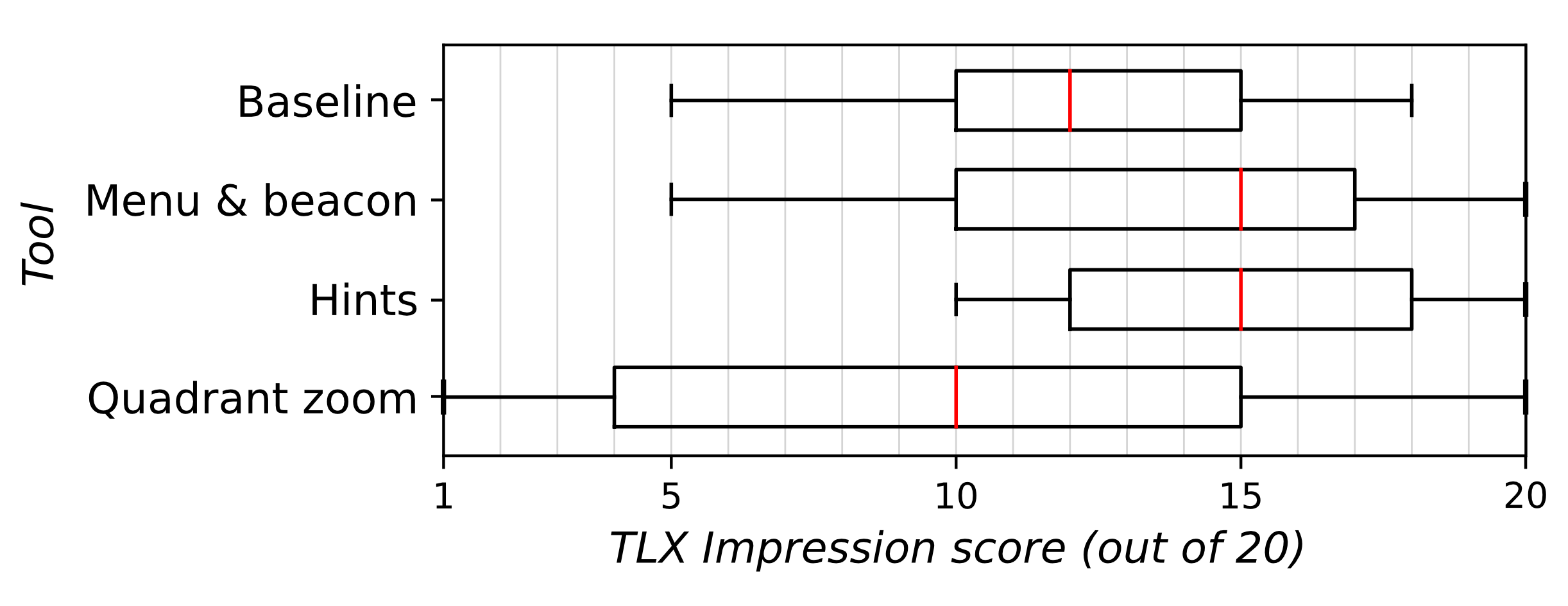}
    \caption{Box plot showing TLX-style impression scores for all four tools tested. Red bars indicate median scores. Participants were polarized on their impression of the zoom tool.}
    \Description{Box plot of TLX-style impression scores on a scale of 1 to 20. The quadrant zoom tool's impression scores span the entire range of possible scores with a median of 10. The zoom tool's median of 10 is the lowest out of all four tools we tested.}
    \label{fig:TLXImpressionZoom}
\end{figure}

The quadrant zoom tool generally evoked mixed feedback from participants. Figure \ref{fig:TLXImpressionZoom} shows a box plot of TLX-style impression scores given by participants for all four tools. The figure shows that participants’ sentiments for the zoom tool spanned the entire range of possible response values, and the median impression was right in the middle of the range (with a value of 10).

Some participants found the quadrant zoom tool to be helpful in surveying small areas, especially since they were surveying the images on smaller screens. The most vocal proponent was P5, who used an Apple iPhone SE --- which has a 4.7-inch/11.9-centimeter display, making it the smallest phone used in our study. P5 said the following after exploring the floor plan with the quadrant zoom tool:

\begin{quote}
    \textit{“The [quadrant zoom tool] gives me an idea of how things are actually in relation to each other. Because when you're moving with your finger and there's small areas, it's kind of hard to tell, like, I didn't know that the house was connected to the balcony [in the floor plan]. And it helped me understand the way that the kitchen area kind of connected to things as well.”} --- \textbf{P5}
\end{quote}

Even some participants who did not have such small devices (P2, P7, and P8) thought that the tool could be useful in situations where areas and sub-areas within the image were very crowded together and otherwise difficult to pinpoint.



Participants' sentiments also pointed to an important insight we encountered during our exploration with blind-accessible zooming and panning (from Section \ref{subsec:tools-zoom}). When it comes to zooming, BLV users find it more important to have a very obvious frame of reference (as is afforded by our quadrant zoom system) over having precise control (as was afforded by our initial zooming and panning system). Zooming systems for BLV users should, thus, emphasize extremely quick and understandable frames of reference rather than levels of control.

Some participants (P1, P4, and P9), however, found the quadrant zoom tool difficult to use by itself. They remarked that, effectively, zooming presented them with a "new image" which happened to be a portion of the main image. If they did not use the menu, beacon, or hints, participants did not feel they could grasp what they were looking at in this "new image," making surveying difficult for them and bringing back some of the issues they found in the baseline image exploration tool. This sentiment indicates that, for many users, the quadrant zoom tool would only be helpful if it came alongside other tools for improving and scaffolding image exploration.


\subsection{ImageAssist vs. Baseline}
\label{subsec:results-general}

\begin{figure}
    \centering
    \includegraphics[width = 0.45\textwidth]{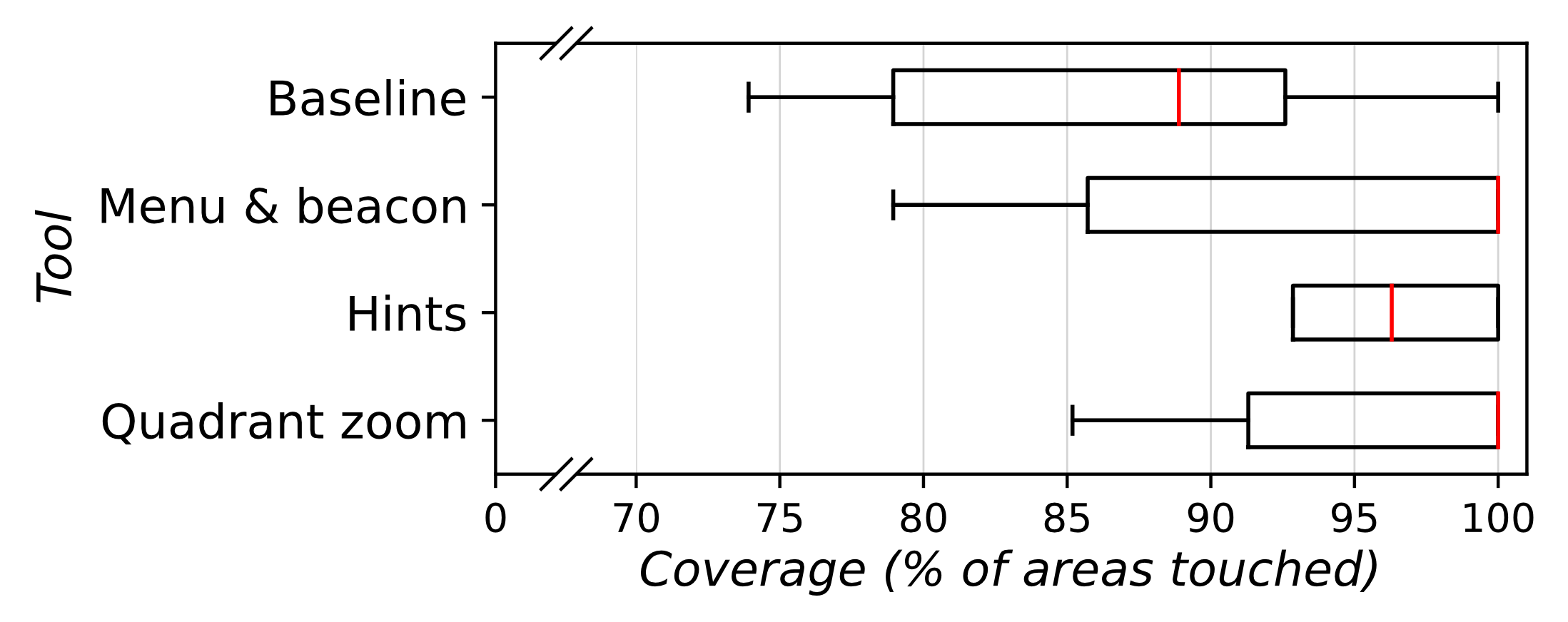}
    \caption{Box plot showing the coverage of participants' touches (i.e., percentage of total areas participants touched while using each tool). Red bars indicate median percentages. Recall that participants could cease exploration whenever they wanted. The baseline tool had a lower median coverage and wider range than any of ImageAssist's component tools, which means that it gives users the least amount of guarantee that they will be able to achieve full coverage.}
    \Description{Box plot of percentages starting (after a break) at 70\% and ending at 100\%. The baseline tool has the lowest median percentage of areas touched (with 89\%) compared to the menu and beacon (100\%), hints (96\%), and quadrant zoom condition (100\%).}
    \label{fig:percent-areas-touched}
\end{figure}

\begin{figure}
    \centering
    \includegraphics[width = 0.45\textwidth]{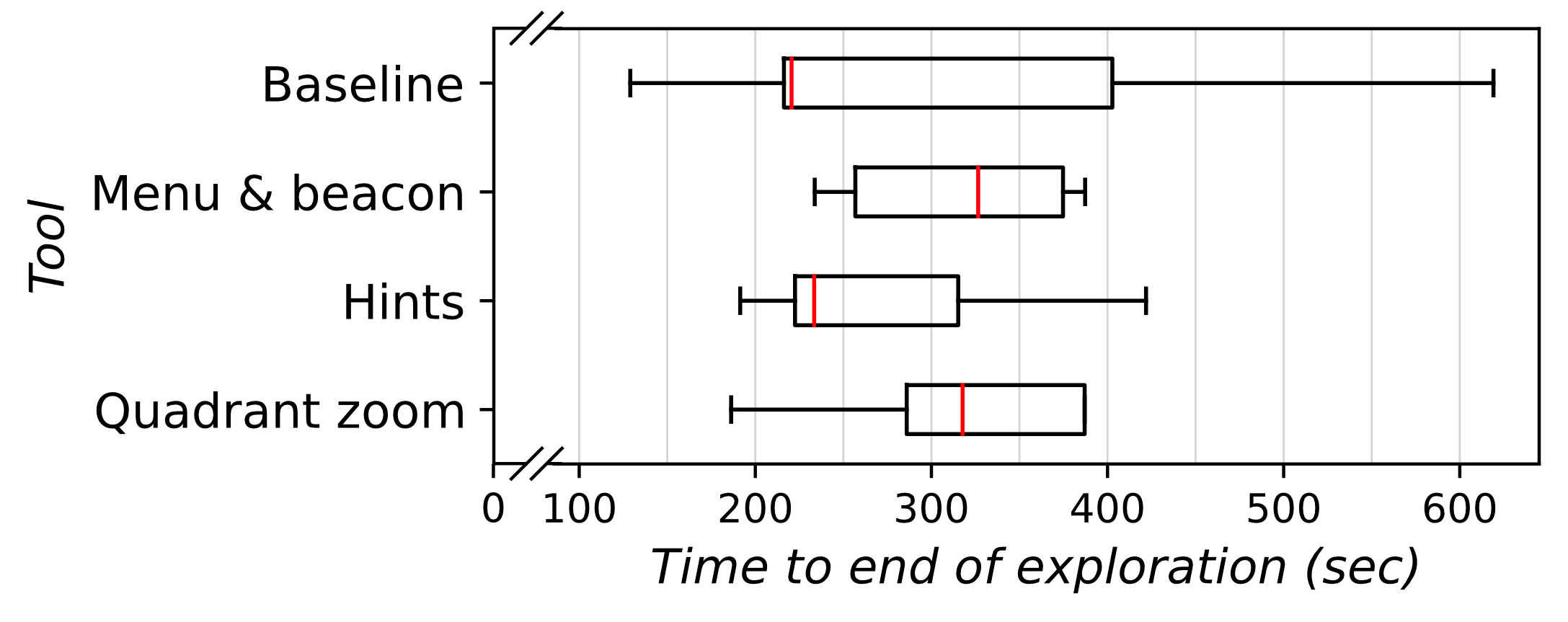}
    \caption{Box plot showing how long (in seconds) participants explored images with a given tool before they voluntarily ceased exploration. Red lines indicate median times. Participants' exploration times were highly variable with the baseline tool when compared with any of ImageAssist's component tools. Many participants found exploring with the baseline tool to be tedious: Some quit early, while others stuck with it for much longer.}
    \Description{Box plots of times starting (after a break) at 100 seconds and ending at 650 seconds. The baseline tool has the highest range of exploration times --- ranging from 160 seconds to 620 seconds. Menu and beacon tool ranges from 290 to 390 seconds; hints tool range from 190 to 420 seconds; and the quadrant zoom tool ranges from 190 to 390 seconds.}
    \label{fig:time-to-end}
\end{figure}

ImageAssist allowed participants to be more sure of themselves when exploring the image --- allowing them to more completely and efficiently survey the image when compared to baseline touchscreen-based image exploration systems.

Figure~\ref{fig:percent-areas-touched} shows participants' average coverage of images using each tool, with coverage defined as the percentage of total areas that participants touched (both top-level areas and sub-areas). If an image had seven areas and the participant touched all seven at some point, for example, their coverage is 100\%. Recall from our study procedure that users could end image exploration at any time --- even if they had not "seen" (i.e., touched) all of the areas within the image.

Ideally, if users are interested in exploring a particular image, their image exploration tool should make it easy for them to explore it with 100\% coverage. In Figure~\ref{fig:percent-areas-touched}, we see that the baseline tool has a lower median coverage score than any of ImageAssist's component tools. It also has the widest range of any of the tools, which means that it gives users the least amount of guarantee that they will be able to achieve full coverage.


Figure~\ref{fig:time-to-end} shows the time that users spent exploring with the four tools before they voluntarily quit. Participants' exploration times were highly variable with the baseline tool, with some participants exploring about five times as long as others. Many participants complained that exploring with the baseline tool was very tedious; some quit early, while others stuck with it for much longer:

\begin{quote}
    \textit{"There's no clear structure as to where I should be looking. [...] I'm done exploring this image. I don't even know if I've seen everything, and honestly, I don't care [to look] anymore."} --- \textbf{P4}
\end{quote}

The three ImageAssist tools have smaller time ranges than the baseline tool, offering users a stronger guarantee that their exploration process will not devolve into a ``worst-case'' scenario.

%% file: sec07-discussion.tex
Our study findings reveal several implications for future image exploration tools for BLV users. Here, we reflect on the findings and propose ideas for next steps.

\subsection{The importance of overviews}

In our main study, we observed that participants highly valued using the menu to get an idea of what was in an image before --- and after --- exploring it, and that having the menu announce all of the elements in the image did not at all discourage them from surveying the image by touch. Indeed, participants who encountered the baseline condition after trying the menu as part of our counterbalancing scheme often complained about missing the overview provided by the menu. This was in spite of us providing a caption --- another type of overview --- for the image before they started exploring the image in every condition.

These observations support the idea that an overview of elements within the image should be an essential part of a touchscreen-based image exploration system. Previous research~\cite{Lee2022a} has shown that such an overview by itself does not facilitate image understanding well but that BLV users may potentially desire using it in conjunction with exploration via touch. Our results confirm the usefulness of such a combination system.

\subsection{What should constitute a hint?}

Our results showed that our BLV participants received the hints tool very positively because it gave them a sense of direction in their exploration. However, participants’ behaviors also indicated that they did not base their understanding of the image solely on those hints. In other words, they began forming their own interpretations of the image, and our conversations with our participants yielded that this ability is something they value but do not experience with alt text and other current mainstream tools designed to help BLV users ``view’’ images.

Granting the ability for BLV users to interpret images also raises questions about what should constitute a “hint.” Our prominence hints, for example, were the outputs of a neural network, and at the end of our study sessions, many participants were curious to learn how exactly we generated these hints. Aside from prominence --- which participants found helpful --- some participants suggested perspective and depth information as possible hints that could be provided by the tool. Future work should investigate what types of information about an image may help BLV users better understand and interpret it. Such information could be well-suited to serve as a hint in an image exploration system.

\subsection{Maximizing both agency and ease of use}


Maximizing \textit{both} agency \textit{and} ease of use is an important design goal accessible tool designers often pursue~\cite{Nair2022}. Our work centers around how we can achieve this within the context of image understanding for BLV users.

Prior work has identified a dichotomy in the ease of use and agency granted by alt text and image exploration systems~\cite{Lee2022a, Morris2018}. Touchscreen-based image exploration systems give BLV users a high degree of agency and control yet are often tedious to use. On the other hand, alt text is very easy to use but does not grant users much agency and control in "seeing" the image. This dichotomy yields two strategies for maximizing \textit{both} agency and ease of use.

One strategy is to start with a technique that is already easy to use even if it does not grant much agency --- such as alt text --- and then imbue more user control within that system. One example of this strategy is Morris et al.'s "progressive detail" alt text~\cite{Morris2018}. It allows BLV users to select how much detail they want to hear when listening to alt text, thus granting them more agency than if they were listening to standard alt text.

Another strategy is to start with a technique that already grants users a high degree of agency and control even if it is not very easy to use --- such as touchscreen-based image exploration systems. The goal, then, becomes smoothing out the user experience so that it becomes easier to use. This paper represents this latter approach, where we introduce a suite of tools so that BLV users are no longer "shooting in the dark" and more easily able to explore images.

We consider both strategies to be valid, however. Future work should continue to investigate extending techniques to grant BLV users a high degree of agency when learning about images while also making them easy to use.

%% file: sec08-applications.tex
Participants' sentiments during our studies revealed that there are times they would want to explore images further via touch. Thus, enhanced touchscreen-based exploration systems can foster an improved understanding of digital images for BLV users in these domains.

\begin{figure*}
    \centering
    \begin{subfigure}{.45\textwidth}
      \centering
      \includegraphics[width=.98\linewidth]{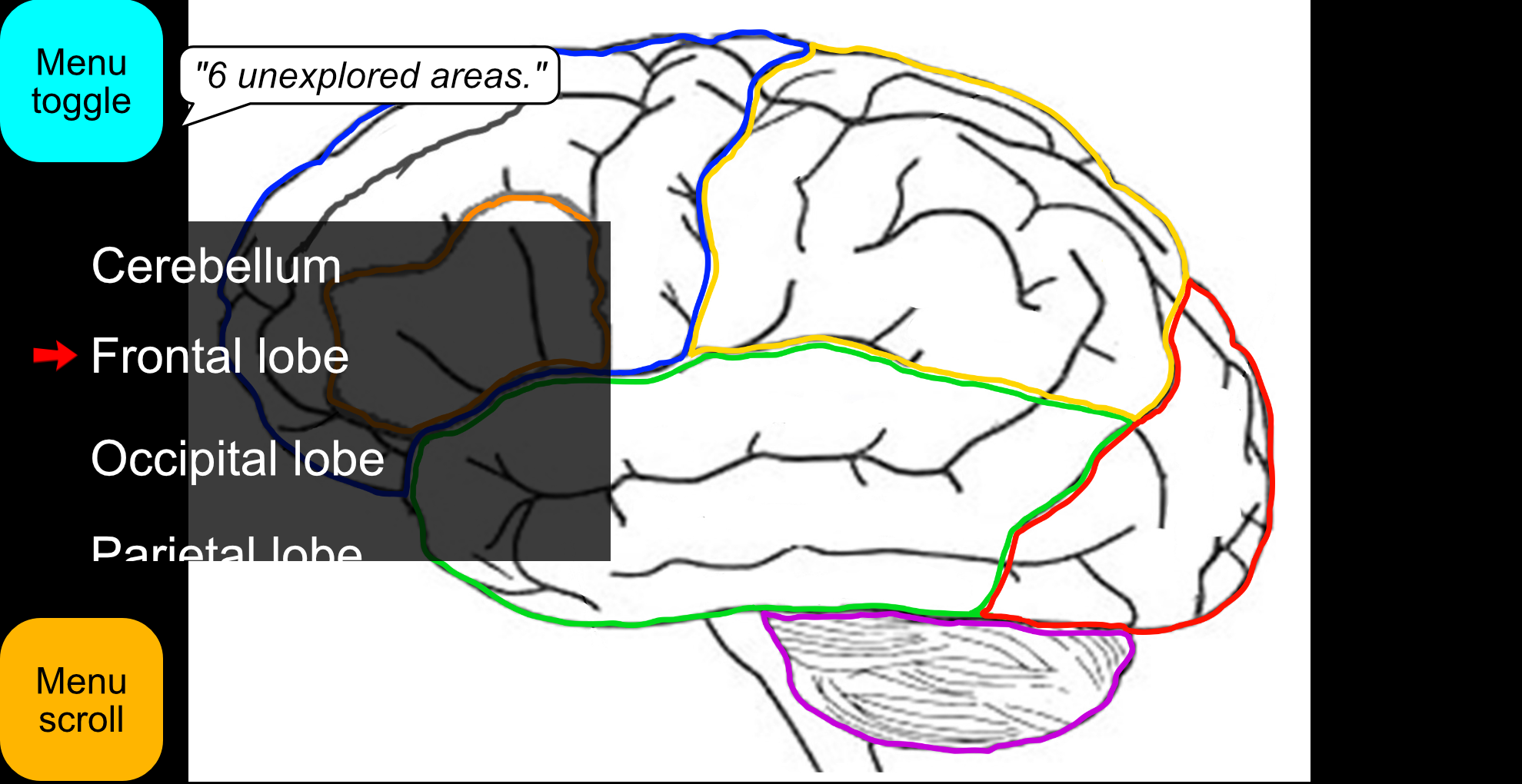}
      \caption{}
    \end{subfigure}%
    \begin{subfigure}{.45\textwidth}
      \centering
      \includegraphics[width=.98\linewidth]{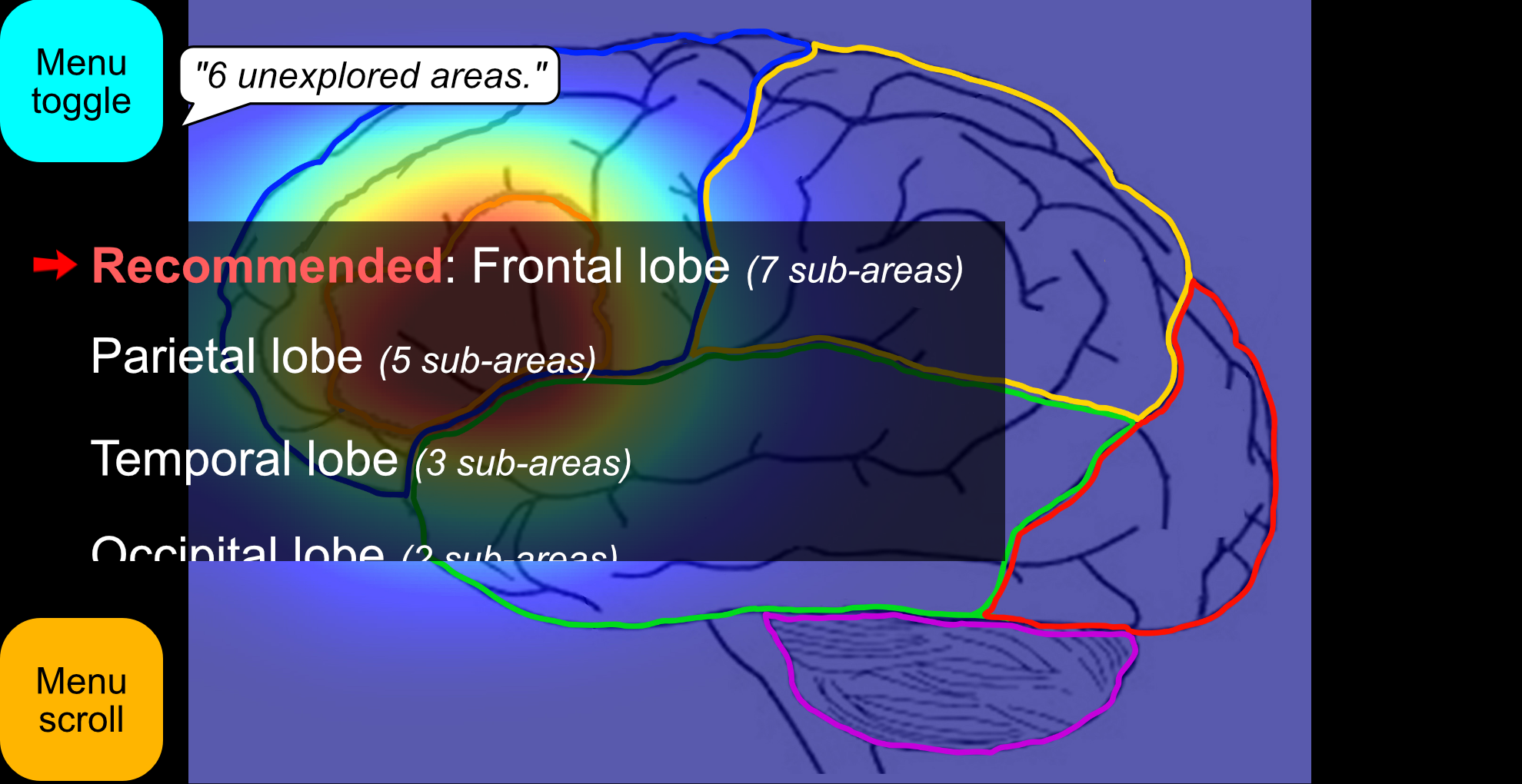}
      \caption{}
    \end{subfigure}
    \caption{ImageAssist in an educational context. (a) Illustration of the menu \& beacon tool on a diagram of the human brain. The regions listed in the menu are the various lobes of the brain. (b) Illustration of the hints tool on a diagram of the human brain. The brightest region of the heatmap represents the activation of the inferior frontal gyrus, which is responsible for speech and language processing in the brain. The inferior frontal gyrus is located in the frontal lobe, which is recommended in the menu and will be announced as the loudest out of all lobes when the user surveys it by touch. The sub-areas within each lobe represent the various gyri (or ridges) within each lobe.}
    \Description{Two panel figure showing ImageAssist's menu and beacon tool and hints tool on a drawing of the human brain looking at the brain from the side. The various lobes of the brain are listed in the menu. For the hints tool, the frontal lobe is recommended and is highlighted on a heatmap superimposed on the diagram.}
    \label{fig:apps-brain}
\end{figure*}

\subsection{STEM Education}

Tactile images play a crucial role in allowing BLV students to understand educational concepts and keep up with their sighted peers in terms of academic performance~\cite{Hasper2015, Stone2019}. With the increasing availability of touchscreen devices, tools like ImageAssist could make scientific diagrams more accessible and informative to BLV STEM students. Indeed, some of our study participants were vocal about how a system like ImageAssist could be leveraged by teachers of students with vision impairments (TVI) to provide a better understanding of concepts:

\begin{quote}
    \textit{"I could see this sort of tool being used by educators. [...] I could see a TVI taking a diagram, labeling it --- using a simple tool --- and sending it to the blind student so they can feel it on their tablet. [...] More students are using tablets, and even blind students are using tablets over something like braille, so I can see something like this being used a lot. The educational implications are staggering."} --- \textbf{P1}
\end{quote}

Figure \ref{fig:apps-brain} shows such an example. In this diagram of a human brain, ImageAssist's menu and beacon tool could give students an overview of the various lobes and ridges of the brain, while ImageAssist's hints tool could be used to emphasize areas of the brain that are activated during a certain state. One way to allow TVIs to more easily annotate these images may involve the use of crowdsourcing --- a technique that has been investigated before in STEM contexts~\cite{Morash2015}.

\begin{figure*}
    \centering
    \begin{subfigure}{.45\textwidth}
      \centering
      \includegraphics[width=.98\linewidth]{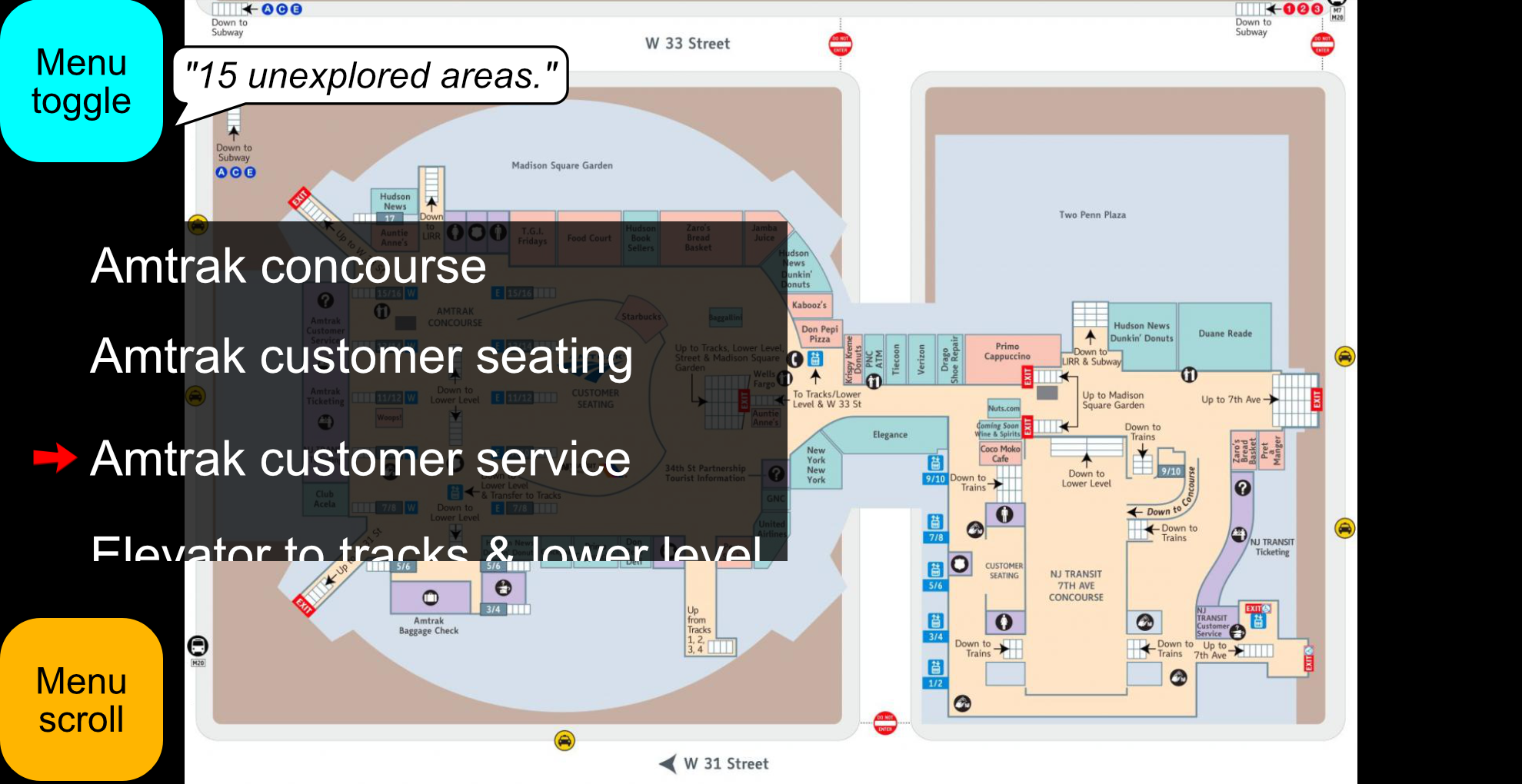}
      \caption{}
    \end{subfigure}%
    \begin{subfigure}{.45\textwidth}
      \centering
      \includegraphics[width=.98\linewidth]{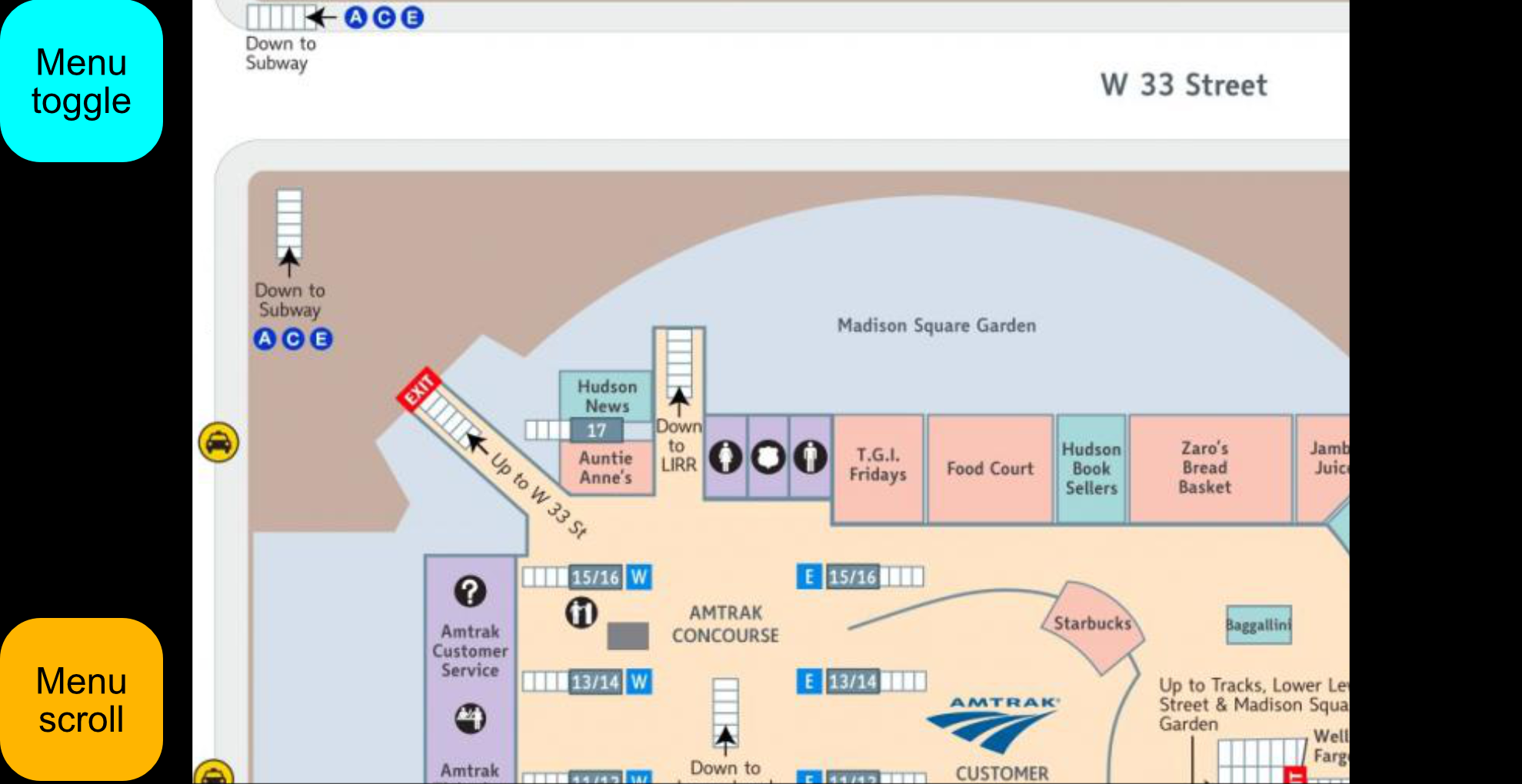}
      \caption{}
    \end{subfigure}
    \caption{ImageAssist in a navigation context. (a) Illustration of the menu \& beacon tool on a map of the upper level of Penn Station in New York City. The regions listed in the menu are the various areas within the map. (b) Illustration of the zoom tool on a map of the upper level of Penn Station in New York City. The user has zoomed into the upper-left quadrant of the map to survey it more closely.}
    \Description{Two panel figure showing ImageAssist's menu and beacon tool and quadrant zoom tool on a map of the upper level of Penn Station, a train station in New York City and the busiest transportation facility in the Western Hemisphere. The various areas shown on the map are listed in the menu, and in the second panel, the user has zoomed into the upper-left corner of the map.}
    \label{fig:apps-penn}
\end{figure*}

\subsection{Map Exploration}

ImageAssist can also make the process of exploring maps of physical world environments within a "rehearsal" context easier --- that is, by allowing BLV users to study maps of areas in-depth prior to physically navigating through them. Prior work has found ``rehearsal'' to increase BLV individuals' independence while navigating~\cite{Ivanchev2014, Meneghetti2012}; however, this process is difficult on smaller screens~\cite{Guerreiro2017}. Participants were enthusiastic about this idea. One example is P5, who stated the following after using the quadrant zoom tool on the floor plan in the main study:

\begin{quote}
    \textit{"You know what would be really cool? Using this to look at a new place before you go there. Basically like how I looked more closely at the rooms [in the floor plan], it would be cool to put in maps of other real places and buildings and stuff like that, so I can get an idea of what the place looks like."} --- \textbf{P5}
\end{quote}

Figure \ref{fig:apps-penn} shows how ImageAssist can help here. A BLV user could use the menu \& beacon to gain an overview of the areas within the environment (here, a complex transportation hub), and then use the zoom tool to survey specific areas in greater detail. These current features can give users a basic understanding of what is on the map and where various points-of-interest are. However, users' understanding of the map can be further enhanced by optimizing ImageAssist specifically for viewing maps --- that is, by explicitly communicating the size, shape, and distance between parts of the map.

\subsection{Social Media and Images of Personal Value to Users}

Prior work has identified that BLV individuals use social media just as much as sighted individuals~\cite{Wu2014, Morris2016}. As visual media, such as images, become dominant, many social media platforms have begun adding features to make images more accessible to BLV users via alt text and descriptions~\cite{Libaw2018}. Tools like ImageAssist, however, can allow BLV users to even more closely explore and learn about images they come across on social media, especially photos that may have personal value. Indeed, some of our BLV participants recounted times when they appreciated having the ability to deeply explore images with existing tools such as Seeing AI~\cite{Microsoft2017}:

\begin{quote}
    \textit{"There was an image --- I was with my family. We went to New York, and we took a picture of the Brooklyn Bridge together. And then I explored the image [using Seeing AI] and I found out there were buildings behind us --- and the sky, clouds, and things like that. So I really liked having that extra info."} --- \textbf{P2}
\end{quote}

\begin{figure*}
    \centering
    \begin{subfigure}{.45\textwidth}
      \centering
      \includegraphics[width=.98\linewidth]{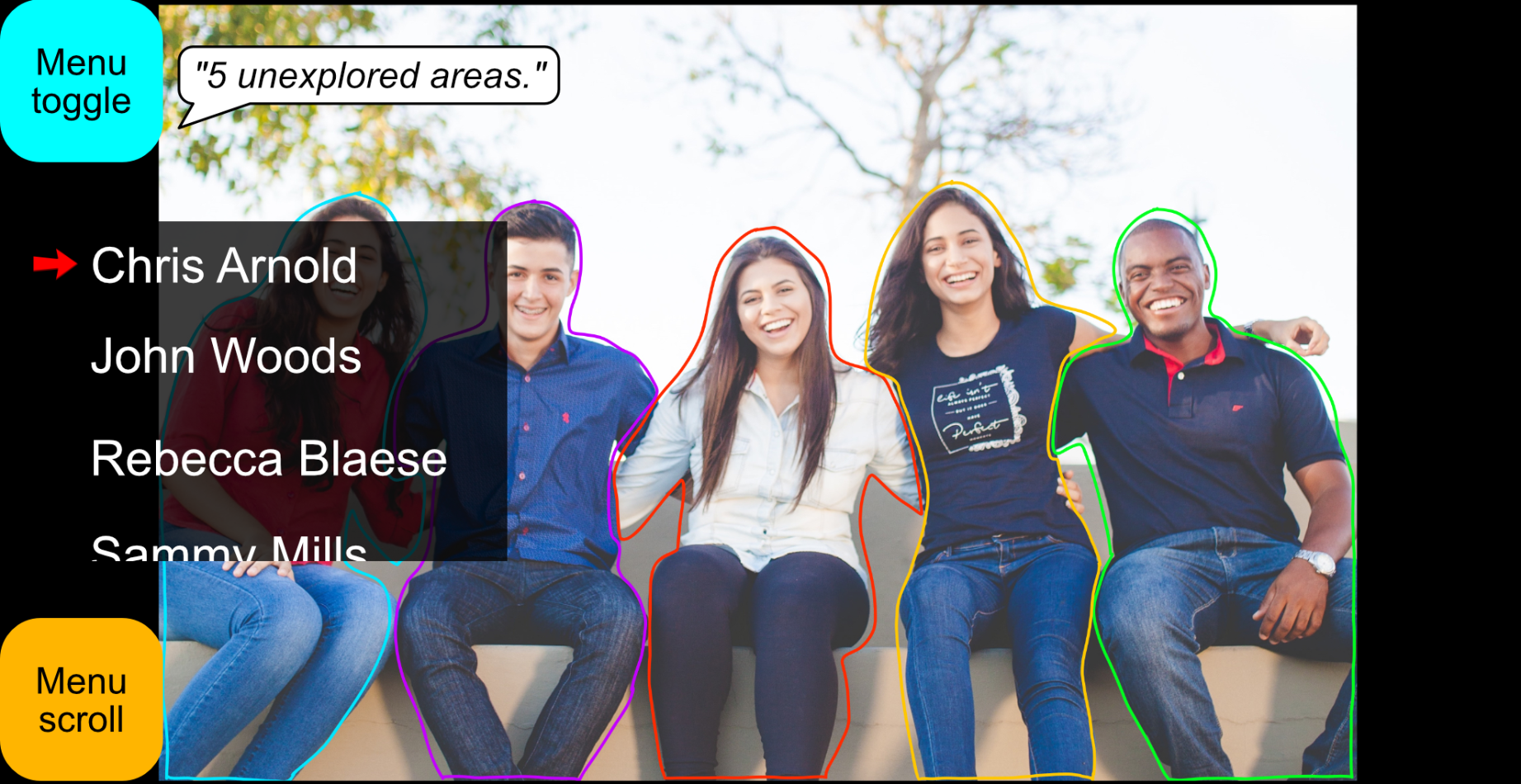}
      \caption{}
    \end{subfigure}%
    \begin{subfigure}{.45\textwidth}
      \centering
      \includegraphics[width=.98\linewidth]{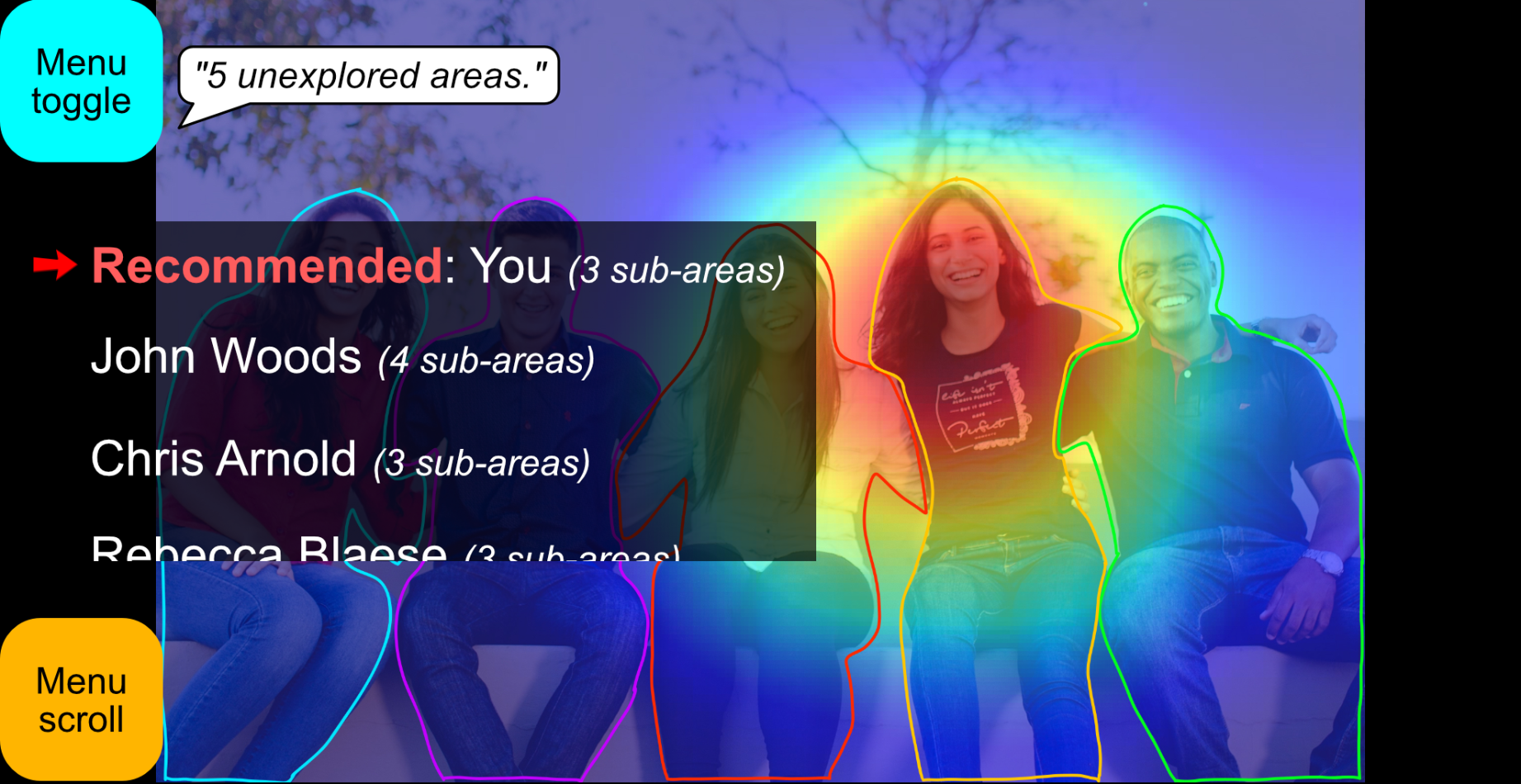}
      \caption{}
    \end{subfigure}
    \caption{ImageAssist in a social media context. (a) Illustration of the menu \& beacon tool on a group photo. The regions listed in the menu are the people shown in the image. (b) Illustration of the hints tool on a group photo. The user is the second person from the right, the brightest region on the heatmap, and recommended in the menu --- this is so that users can find themselves quickly. The sub-areas under each person represent aspects of their appearance (e.g., hair, clothing, and accessories), which users can survey by entering the area represented by the person and surveying via touch and/or the menu \& beacon.}
    \Description{Two panel figure showing ImageAssist's menu and beacon tool and hints tool on a group photo of five smiling people sitting side-by-side with their arms over each other's shoulders and looking at the camera. The people's names are listed out in the menu. For the hints tool, the user (in this case, labeled in the menu as "You") is recommended in the menu and is highlighted on a heatmap superimposed on the photo.}
    \label{fig:apps-sm}
\end{figure*}

Figure~\ref{fig:apps-sm} shows an example of how ImageAssist could work on a group photo posted to social media. A BLV user can use the menu to gain an overview of who is in the image and any objects of interest and the beacon to find any people or elements they are interested in. The hints tool can additionally streamline the process of helping users find \textit{themselves} or people they are close to within the photo.

%% file: sec09-limitations.tex
While our work proposes solutions for improving touchscreen-based image exploration for BLV users, we acknowledge several limitations.

Our studies involved us manually defining the bounds of areas within our test images. However, some existing touchscreen-based image exploration systems such as ImageExplorer~\cite{Lee2022a} and Seeing AI~\cite{Microsoft2017} make use of automated, machine learning-based techniques to segment the image (and define these bounds). ImageAssist operates independently of the exact method used to define area bounds within an image; however, we acknowledge that testing ImageAssist with images segmented \textit{automatically} could yield additional insights into how segmentation method and quality could affect users' impressions.

Our study sample included only nine BLV individuals, and their preferences for image exploration tools may not necessarily be representative of the blind and low vision community at large. In particular, we created ImageAssist's features in response to the themes that arose in our formative study, and there may be other possible themes that we did not consider in our current exploration. There may also be other possible \textit{designs} that we did not consider either. For example, although we tested quadrant zooming for design goal G3, there may be other forms of zooming --- such as by fitting a top-level region to the screen --- that could also yield interesting insights into user behaviors. Finally, more research is needed to adapt the tools we created for BLV users with other disabilities --- such as severe hearing and motor impairments --- that may impede effective use of our tools.

%% file: sec10-conclusion.tex
In this work, we designed and evaluated ImageAssist, a set of three tools intended to scaffold the image exploration process so that BLV users can better take advantage of such systems' benefits when exploring digital images. Through a formative study, we had BLV users provide their feedback on a state-of-the-art exploration system and found important issues that needed to be addressed by ImageAssist. We designed and implemented three tools in response to these problems and tested them with BLV users in a user study to see how well they addressed the issues we identified.

In the formative study, we found that participants scanned images in an inefficient manner, missed image elements even when trying to find everything, and faced difficulties in surveying small and tightly clustered elements. In response to these issues, we developed ImageAssist, a set of three tools that supplement state-of-the-art touchscreen-based image exploration: a menu \& beacon tool, a hints tool, and a quadrant zoom tool. We tested these tools in a user study and found that participants tended to use the menu \& beacon as an overview of the image at the beginning and end of exploration; that they used the hints to form their own interpretations of the image while exploring; and that they had mixed feelings about the quadrant zoom tool possibly being helpful, but also potentially complicating exploration.

We hope that creating tools to improve touchscreen-based image exploration systems can make these systems more open and accessible to BLV users, allowing them to better understand images that they come across in their everyday lives.